\documentclass[prd,appendix,amsmath,amssymb,amsfonts,graphicx,epsfig,10pt,twocolumn]{revtex4}

\usepackage{amssymb,amsmath,amsthm,graphicx,amscd,xcolor}
\def\E{\mathbf{E}}

\def\B{\mathbf{B}}
\def\D{\mathbf{D}}
\def\H{\mathbf{H}}

\def\x{\mathbf{x}}
\def\k{\mathbf{k}}

\def\E{\mathbf{E}}
\def\A{\mathbf{A}}

\def\B{\mathbf{B}}
\def\D{\mathbf{D}}
\def\H{\mathbf{H}}

\def\x{\mathbf{x}}
\def\k{\mathbf{k}}

\allowdisplaybreaks

\usepackage{amsfonts}
\usepackage{amssymb}
\usepackage{amsmath}
\usepackage{MnSymbol}
\usepackage{graphicx}
\date{\today}

\begin{document}

\title{Analogues of gravity-induced instabilities in anisotropic metamaterials}

\author{Caio C.\ Holanda \surname{Ribeiro}}
\email{caiocesarribeiro@ifsc.usp.br}
\affiliation{Instituto de F\'\i sica de S\~ao Carlos,
Universidade de S\~ao Paulo, Caixa Postal 369, 13560-970, 
S\~ao Carlos, S\~ao Paulo, Brazil}

\author{Daniel A.\ Turolla \surname{Vanzella}}
\email{vanzella@ifsc.usp.br}
\affiliation{Instituto de F\'\i sica de S\~ao Carlos,
Universidade de S\~ao Paulo, Caixa Postal 369, 13560-970, 
S\~ao Carlos, S\~ao Paulo, Brazil}

\begin{abstract}

In the context of field theory in curved spacetimes, 
it is known that suitable background spacetime geometries can trigger 
instabilities of fields, leading to exponential growth of their (quantum and classical) fluctuations --- a phenomenon called {\it vacuum awakening} in the quantum context,
which in some classical scenarios seeds {\it spontaneous scalarization/vectorization}. 
Despite its conceptual interest, an actual observation in nature of this effect is uncertain since it depends on the existence of fields with appropriate masses and couplings
in strong-gravity regimes. Here, we  propose analogues for this gravity-induced instability
based on nonlinear optics  of metamaterials which could, in principle, be observed in laboratory.

\end{abstract}

\maketitle

\section{Introduction}

The influence of a background material medium on the propagation of mechanic and electromagnetic waves is well known to be 
formally analogous to that of an effective curved spacetime geometry. This idea was first presented,
in the electromagnetic/optical context, by 
Gordon in 1923~\cite{Gordon1923} and it has since been
developed in a number of different scenarios, particularly after Unruh's~\cite{Unruh81} and 
Visser's~\cite{Visser1998} works on 
acoustic analogues of black holes and their associated Hawking-like radiation. More recent applications of this formal analogy include mimicking in material media
quantum lightcone fluctuations~\cite{Bessa2016} and anisotropy in cosmological spacetimes~\cite{Visser2007}.
The most appealing feature of these condensed-matter analogues of gravitational
backgrounds is the possibility of observing in laboratory subtle but conceptually interesting effects 
which can be virtually unobservable in their original contexts ---  Hawking radiation being certainly the most emblematic among them,
with claims of having already been observed in laboratory~\cite{Wein11,Unruh14,Nova2019}.

An interesting effect in the context of (quantum) fields in curved spacetimes is the triggering of field instabilities 
due to the background spacetime geometry ---
a phenomenon called vacuum awakening in the quantum context~\cite{LV,LMV,LMMV,MMV}. 
These gravity-induced instabilities exponentially amplify vacuum fluctuations to the point they decohere and seed classical perturbations~\cite{LLMV}, which, depending on field parameters,
eventually evolve to a nonzero classical
field configuration (``spontaneous scalarization'' in the case of scalar fields~\cite{PCBRS,DE1,DE2,B}), stabilizing the whole system.
More recently, this mechanism was also predicted to occur for massless spin-$1$ fields through appropriate 
nonminimal couplings~\cite{Cardoso2019} and, in analogy with the scalar case, the stabilization process was termed ``spontaneous vectorization.'' To the best of our knowledge, condensed-matter and optical analogues of these 
gravity-induced
instabilities have not been proposed to this date.
In this work, we propose and explore possible analogues of gravity-induced instabilities in the context of
electromagnetism in  polarizable/magnetizable anisotropic (meta)materials. 

Electromagnetic instabilities in flat 
spacetime are expected to occur in some
materials.
 One celebrated example appeared in the context of plasma physics in the late 1950s and became known as
  Weibel instability~\cite{Weibel1959}. The system, a neutral plasma whose components have anisotropic velocity 
 distribution, possesses growing  electromagnetic transverse waves. Related effects have been studied 
 since then, with recent applications to solar plasma instability~\cite{Rubab2016} and solid state devices \cite{Silveirinha2017}. Moreover, causal aspects of classical propagation in active materials were 
 discussed in Ref.~\cite{Nistad2008}, 
 where properties of the refractive index were established. 
Nevertheless, besides the fairly recurrence in the literature, usually quantization in such scenarios is not considered \cite{Matloob1995,Matloob1997,Raabe2007} or it is 
 regarded as inconsistent
 \cite{Huttner1992,Raabe2008}.

It is noteworthy that instability of the electromagnetic field is always accompanied by evolution of the background, 
ending with the stabilization of the system as a whole. In the case of gravity-induced instability, the gravitational 
field changes with time, whereas electromagnetic instability in the presence of plasmas involves growing plasmons. 
In the case of electromagnetic fields in the presence of matter, for whatever form of the interaction with the 
background, the field's evolution is ruled by Maxwell's equations in the presence of polarizable/magnetizable 
media, and the interaction with the background is encapsulated in the functional dependence of the electric 
displacement (magnetic) vector field $\D$ ($\H$) with the true (microscopic) fields $\E$ and $\B$. If the magnitudes involved are 
small (e.g., in the beginning of the instability action), these functional relations become linear and 
one may find the form of the coefficients for such systems. For the case of Weibel instability, for instance, if the 
velocity anisotropy is taken in the $z$ direction, the instability is modelled by a negative squared refractive index in 
the direction perpendicular to $z$.

We 
apply Gordon's method to propose a family of optical-based analogue models for electromagnetic fields presenting 
instabilities in curved spacetimes. We show how anisotropies of the background enter the effective equations in the form of nonminimal couplings, and in the case of strong anisotropy (just like for the Weibel instability), this coupling results in unstable solutions. 
We also discuss that for these systems the stabilization process occurs through the nonlinear nature of the background, which may seed spontaneous vectorization in analogy to the Einstein's field equations in the gravitational scenario.  

The paper is organized as follows. In Sec.~\ref{sec:cov}, we present the covariant formalism of electromagnetism
in anisotropic polarizable/magnetizable materials, establishing the formal analogy with nonminimally-coupled 
electromagnetism in curved spacetimes. In Subsec.~\ref{subsec:QED}, we consider a particular type of nonminimal
coupling inspired by one-loop quantum electrodynamics (QED) corrections to electromagnetism in
curved spacetimes.
In Sec.~\ref{sec:ani}, we apply the formalism presented in the previous
section to the scenario of a plane-symmetric anisotropic medium at rest in an inertial frame. Although 
plane-symmetric curved spacetimes (in four dimensions) are not really (physically) appealing, we consider
this scenario for its simplicity and for its possible implications for the physics of the material medium.
We  construct the electromagnetic quantum-field operator $\hat{{\bf A}}$ 
(in the generalized Coulomb gauge) in the standard-vacuum representation, discuss
the conditions for appearance of instabilities and their types (Subsec.~\ref{subsec:insta}), and
present a concrete example (homogeneous medium; Subsec.~\ref{subsec:explane})
where calculations can be carried over to the end. In Sec.~\ref{sec:spherical}, we
repeat the treatment of the previous section, but now for a more appealing scenario
on the gravitational side: spherically-symmetric,
stationary anisotropic media. Conditions for triggering instabilities and their types are shown to be
very similar to those in the plane-symmetric case (Subsec.~\ref{subsec:instass}). As a concrete application,
in Subsec.~\ref{subsec:exss} we show how to mimic QED-inspired nonminimally-coupled electromagnetism
 in the background spacetime of a Schwarzschild black hole. Then, Sec.~\ref{sec:stabilization} is dedicated to 
 discuss possible stabilization mechanisms which might bear analogy to some curved-spacetime
 phenomena, such as spontaneous vectorization~\cite{Cardoso2019} and particle 
 bursts due to tachyonic 
 instability~\cite{LLMV2}.
 Finally, in Sec.~\ref{sec:final} we present some final remarks. We leave for an appendix 
 tedious calculations related to the orthonormalization of modes of Sec.~\ref{sec:spherical}.
 We adopt the abstract-index notation to represent tensorial quantities
 (see, e.g., Ref.~\cite{Wald}) and, unless stated otherwise, we use natural units (in which $\hbar = c = 1$).

\section{Covariant electromagnetism in anisotropic material media}
\label{sec:cov}

Electromagnetism in material media, in flat spacetime
and in 
the absence of free charges, 
is described  by two antisymmetric (observer-independent) tensors, $F_{ab}$ and $G^{ab}$, 
satisfying the macroscopic covariant Maxwell's equations,
\begin{eqnarray}
\partial_a G^{ab}&=&0,\label{mx1}\\
\partial_{[a}F_{bc]}&=&0,\label{mx2}
\end{eqnarray}
where $\partial_a$ is the derivative operator compatible with the flat metric $\eta_{ab}$ (but in arbitrary coordinates) and
the square brackets denote antisymmetrization over the indices enclosed by them. 
These equations must be supplemented by
medium-dependent constitutive relations between $F_{ab}$ and $G^{ab}$, as well as initial and boundary conditions, in order
to provide a well-posed problem. These constitutive relations are usually set at the level of (observer-dependent)  fields
$E_a$, $B^a$, $D^a$, and $H_a$, related to $F_{ab}$ and $G^{ab}$ through
\begin{eqnarray}
E_a&=&F_{ab}u^b,\label{ell}\\
D^a&=&G^{ab}u_b,\label{dil}\\ 
B^a&=&-\frac{1}{2}\epsilon^{abcd}F_{bc}u_d,\label{ibl}\\
H_a&=&-\frac{1}{2}\epsilon_{abcd}G^{bc}u^d,\label{mgl}
\end{eqnarray} 
where $u^a$ is the four-velocity of the observer measuring these fields and $\epsilon_{abcd}$ is the Levi-Civita
pseudo-tensor (with $\epsilon_{0123} = \sqrt{-\eta}$, $\eta := \det(\eta_{\mu\nu})$). Moreover, the constitutive relations
usually take a simpler form in the reference frame in which the medium is (locally and instantaneously) at rest. 

Here, we consider a polarizable and magnetizable medium whose constitutive relations in its  instantaneous rest frame take the form
\begin{eqnarray}
D^a&=&\varepsilon^{ab} E_b, \label{DE}\\ 
H_a&=&\mu_{ab} B^b, \label{HB}
\end{eqnarray}
where the  tensors $\varepsilon^{ab}$
and $\mu_{ab}$ may depend on spacetime coordinates, and the system is assumed dispersionless. We return to this point later.
The fact that Eqs.~(\ref{DE},\ref{HB}) are valid in the medium's instantaneous rest frame means that the 
fields $E_a$, $B^a$, $D^a$, and $H_a$  appearing in them
are related to $F_{ab}$ and $G^{ab}$ through Eqs.~(\ref{ell}-\ref{mgl}) with $u^a = v^a$, the medium's four-velocity field.
We proceed by splitting the ``spatial''~\cite{Foot1}
tensors $\varepsilon^{ab}$ and $\mu_{ab}$ into isotropic and traceless anisotropic parts,
\begin{eqnarray}
\varepsilon^{ab}&=&\varepsilon\,h^{ab}+\chi_{(\varepsilon)}^{ab},\label{epssplit}\\
\mu_{ab}&=&\mu^{-1}h_{ab}+\chi^{(\mu)}_{ab}, \label{musplit}
\end{eqnarray}
where $h^a_{\;b} :=\delta^a_{b} +v^a v_b$ is the projection operator orthogonal to $v^a$. Inverting 
Eqs.~(\ref{dil},\ref{mgl}) (with $u^a = v^a$),
\begin{eqnarray}
G^{ab} = 2 v^{[a} D^{b]}-\epsilon^{abcd}H_{c}v_d,\label{GDH}
\end{eqnarray}
and substituting Eqs.~(\ref{DE}-\ref{musplit}) and (\ref{ell},\ref{ibl}), we obtain
\begin{eqnarray}
G^{ab}
=( g^{ac}g^{bd}+\chi^{abcd})F_{cd},
\label{GF}
\end{eqnarray}
where we have defined the tensors 
\begin{eqnarray}
g^{ab}&:=&\frac{1}{\sqrt{n}}\left[\eta^{ab}-(n^2-1)v^{a}v^b\right],\label{gup}\\
\chi^{abcd}&:=&\left(\frac{n}{\mu}-1\right)g^{a[c}g^{d]b}-2 \chi_{(\varepsilon)}^{\left[a\left|[c\right.\right.}v^{d]} v^{\left|b\right]}\nonumber \\& &
+\frac{1}{2}\epsilon^{abef}\epsilon^{cdgh}\chi^{(\mu)}_{eg}v_f v_h,\;\;\;
\label{chiup}
\end{eqnarray} 
and the squared refractive index $n^2=\mu\varepsilon$.
The idea, then, is to consider the symmetric tensor $g_{ab}$, defined through $g_{ab} g^{bc} = \delta^c_a$, as an {\it effective metric} of a curved background spacetime perceived by the electromagnetic
field $F_{ab}$. Note that the components of $g^{ab}$ and $\eta^{ab}$ satisfy
\begin{equation}
\det\left(g^{\alpha\beta}\right)=\det\left(\eta^{\alpha\beta}\right)\label{detgup}
\end{equation} 
and, thus, $\sqrt{-\eta}=\sqrt{-g}$, where $g:=\det(g_{\alpha\beta})$. One can easily check that  $g_{ab}$ is explicitly given by
\begin{eqnarray}
g_{ab}={\sqrt{n}}\left[\eta_{ab}+\frac{(n^2-1)}{n^2}v_{a}v_b\right].
\label{gdown}
\end{eqnarray} 
Therefore, in an arbitrary coordinate system, Eq.~(\ref{mx1}) reads
\begin{eqnarray}
0&=&\frac{1}{\sqrt{-\eta}}
\partial_\alpha
\left(\sqrt{-\eta}\,
G^{\alpha\beta}\right)
= \frac{1}{\sqrt{-g}}\partial_\alpha
\left(\sqrt{-g}\,
G^{\alpha\beta}\right).
\label{mx1g}
\end{eqnarray}

Up to this point, it was understood that the physical background metric $\eta_{ab}$ and its inverse $\eta^{ab}$ were responsible for lowering and raising tensorial indices.
Now, with the introduction of an effective metric $g_{ab}$, we should be careful when  performing these isomorphisms. In order to minimize chances of
confusion, we shall avoid lowering and raising 
tensorial indices using the effective metric,  making explicit most appearances of $g_{ab}$ and $g^{ab}$
in the equations below, with few  exceptions which will be clearly stated. One obvious exception is 
the definition of $g_{ab}$ as the inverse of $g^{ab}$. Another such exception is
the use of $\nabla_a$ to denote covariant derivative compatible with $g_{ab}$.
With this in mind, from Eqs.~(\ref{mx2}) and (\ref{mx1g}), the electromagnetic tensor $F_{ab}$ satisfies
\begin{eqnarray}
0 &=&\nabla_{[a}F_{bc]}, \label{mx2cov} \\
0&=&
\nabla_a
\left[( g^{ac}g^{bd}+\chi^{abcd})F_{cd}\right].
\label{mx1cov}
\end{eqnarray}
Notice that Eqs.~(\ref{mx2cov}) and (\ref{mx1cov}) applied to homogeneous
($\nabla_a \varepsilon=\nabla_a \mu = 0$), isotropic ($\chi_{(\varepsilon)}^{ab}=0=\chi^{(\mu)}_{ab}$) materials, with {\it arbitrary}  4-velocity field $v^a$, lead to the same equations which
rule minimally-coupled vacuum electromagnetism in a curved spacetime with metric $\sqrt{\mu/n}\,g_{ab}$. Optical analogue models in these configurations with $\mu=1$ were studied in \cite{Ulf1999,Ulf2000}. 
Here, we shall focus on electromagnetism in anisotropic materials, more specifically, materials with only ``shear-like'' anisotropies:
$\chi_{(\varepsilon)}^{[ab]} = 0= \chi^{(\mu)}_ {[ab]}$.
In this case, the 
tensor $\chi^{abcd}$ defined in Eq.~(\ref{chiup}) has the same algebraic symmetries as
the Riemann  curvature  tensor, namely, $\chi^{abcd} = \chi^{cdab}$ and $\chi^{a[bcd]} = 0$ --- in addition to $\chi^{abcd} = \chi^{[ab][cd]}$, which is always true.

The Eqs.~(\ref{mx2cov}) and (\ref{mx1cov}) can be seen as analogous to some nonminimally-coupled
electromagnetic field equations in curved spacetime.
Although in general $\chi^{abcd}$ is independent of  the Riemann tensor associated with the effective metric $g_{ab}$, one can construct
cases where they are related. This is interesting because some one-loop QED corrections to Maxwell's field equations  in curved spacetime~\cite{DH,Lemos}
can be emulated by such  nonminimal coupling, as we shall discuss below,  in Subsec.~\ref{subsec:QED}.

Before considering particular applications of the equations above, let us define a sesquilinear form on the space of complexified
solutions, which will be relevant when applying the canonical quantization procedure.
As usual, let us solve Eq.~(\ref{mx2cov}) by introducing the 4-potential $A_a$ such that 
$F_{ab} = \nabla_{a}A_b-\nabla_bA_a$. Then,
let $F_{ab}$ and $F_{ab}'$ be two {\it complex} solutions of Eq.~(\ref{mx1cov}), associated to $A_a$ and $A_a'$, respectively. With overbars representing complex conjugation, we
contract $\bar{A}_b$ (resp., $A_b'$) with Eq.~(\ref{mx1cov}) applied to $F_{cd}'$ (resp., $\bar{F}_{cd}$)
and subtract one from the other, arriving at
\begin{eqnarray}
\nabla_a\left[\left(g^{ac}g^{bd}+\chi^{abcd}\right)\left(\bar{A}_b F_{cd}'-A_b'\bar{F}_{cd}\right)\right] = 0.
\label{cont}
\end{eqnarray}
This continuity-like equation ensures that the quantity
\begin{eqnarray}
\left(A,A'\right):=i \int_{\Sigma} d\Sigma\,N_a\left(g^{ac}g^{bd}+\chi^{abcd}\right)
\left(\bar{A}_b F_{cd}'-A_b'\bar{F}_{cd}\right)\!\!\!\!\!
\nonumber \\
\!\!\!\!
\label{sesq}
\end{eqnarray}
is independent of the space-like hypersurface $\Sigma$ where the integration is performed --- provided we restrict attention to
solutions satisfying  ``appropriate'' boundary condition ---,
where $d\Sigma$ is the physical volume element on $\Sigma$ and $N_a = \eta_{ab} N^b$, 
with $N^a$ being a unit, future-pointing vector orthogonal to $\Sigma$ (according to $\eta_{ab}$).
More specifically, considering that the system of interest is contained in the spacetime region ${\cal M}\cong T\times \Sigma$,
where $T\subseteq {\mathbb R}$ is a real open interval, then the appropriate boundary condition amounts to imposing that
the flux of the (sesquilinear) current appearing in Eq.~(\ref{cont}) vanishes through $T\times\dot{\Sigma}$ (where
$\dot{\cal S}$ denotes the boundary of the space 
${\cal S}$). In particular, in stationary situations which we shall treat here,
this condition translates to
\begin{eqnarray}
\int_{\dot{\Sigma}}dS\,s_a \left(g^{ac}g^{bd}+\chi^{abcd}\right)\left(\bar{A}_b F_{cd}'-A_b'\bar{F}_{cd}\right)=0,
\label{bcgen}
\end{eqnarray}
where $dS$ is the physical area element on $\dot{\Sigma}$ and $s^a$ is the unit vector field normal to $T\times\dot{\Sigma}$ (according to
$\eta_{ab}$).
Thus, these conditions being satisfied, Eq.~(\ref{sesq}) provides a legitimate sesquilinear form on the space
${\cal S}_{\mathbb C}$ of complex-valued solutions
of Eqs.~(\ref{mx2cov}) and (\ref{mx1cov}). Notice that for pure-gauge solutions --- i.e., 
$A_a = \nabla_a \psi$, for some 
scalar function $\psi$ ---, 
$(A, A) = 0$. (The converse, however, is not true.)

The relevance of this sesquilinear form is that it provides a legitimate
inner product on a (non-unique choice of) subspace ${\cal S}_{\mathbb C}^+\subsetneq {\cal S}_{\mathbb C}$ of 
``positive-norm solutions,'' which, together with its complex conjugate 
${\cal S}_{\mathbb C}^-\subsetneq {\cal S}_{\mathbb C}$, generates all solutions: ${\cal S}_{\mathbb C}$:
${\cal S}_{\mathbb C}^+\oplus{\cal S}_{\mathbb C}^- = {\cal S}_{\mathbb C}$.
Loosely speaking, upon completion, ${\cal S}_{\mathbb C}^+$ yields  
a Hilbert space ${\cal H}$ from which the (symmetrized) 
Fock space ${\cal F}_s({\cal H})$ is
canonically
constructed to represent states of the electromagnetic field. In particular,
choosing ${\cal S}_{\mathbb C}^+$ to be generated by positive-frequency solutions
(those proportional to $e^{-i \omega t}$, with $\omega>0$), the vacuum state of this Fock representation
corresponds to the usual physical vacuum state of the field. 

\subsection{QED-inspired nonminimal couplings}
\label{subsec:QED}

As mentioned earlier, 
 Eqs.~(\ref{mx2cov}) and (\ref{mx1cov}) can be interpreted as ruling electromagnetism
 in curved spacetimes with some QED-inspired nonminimal coupling $\chi^{abcd}$ with the
 background geometry. In fact,  
 in the one-loop-QED approximation~\cite{DH,Lemos},
 \begin{eqnarray}
 \chi^{abcd} = \alpha_1 R^{abcd}+{\alpha_2}R^{[a|[c} g^{d]|b]}
 +{\alpha_3} R\, g^{a[c} g^{d]b},
 \label{QED}
 \end{eqnarray}
 with $\alpha_1 =-\alpha_2/13 = 2\alpha_ 3 = -\alpha /(90 \pi m_e^2)$, where $\alpha$ is the fine-structure
 constant,  $m_e$ is the electron's mass, and $R^{abcd}$, $R^{ab}$, and $R$ are, respectively, the Riemann,
 Ricci and Ricci-scalar curvature tensors associated with the (effective) metric $g_{ab}$. By leaving
 $\alpha_1, \alpha_2,\alpha_3$ unconstrained, Eq.~(\ref{QED}) represents a three-parameter family of 
 couplings of the electromagnetic field 
 with the background effective geometry --- see Ref.~\cite{BL} for some interesting particular cases.
 
 For a generic
 medium, $\chi^{abcd}$ is not related to the geometry associated with $g_{ab}$. 
 However, we can
 simulate couplings given by Eq.~(\ref{QED}) by conveniently relating  $n$ and $v^a$ (which determine $g_{ab}$)
 with $\mu$ and the anisotropic tensors
 $\chi_{(\varepsilon)}^{ab}$ and $\chi^{(\mu)}_{ab}$ (which appear in $\chi^{abcd}$). From 
 Eqs.~(\ref{chiup}) and (\ref{QED}), and their contractions with $g_{ab}$,
\begin{eqnarray}
g_{bd}\chi^{abcd}& =&\frac{3}{2}\left(\frac{n}{\mu}-1\right)g^{ac}+\frac{\chi_{(\varepsilon)}^{ac}}{2n^{3/2}}+
\frac{n^{3/2}}{2}g^{ab}g^{cd}\chi^{(\mu)}_{bd}
\nonumber \\ &=& \left(\alpha_1+\alpha_2/2\right)R^{ac}+(\alpha_2/4+3\alpha_3/2)R\,g^{ac},
\nonumber \\
\label{chiRab}
\\
g_{ac}g_{bd}\chi^{abcd}& =&6\left(\frac{n}{\mu}-1\right) = (\alpha_1+3\alpha_2/2+6\alpha_3)R,
\label{chiRScalar}
\end{eqnarray}
we can solve for $\mu$ and the anisotropic tensors,
obtaining:
\begin{eqnarray}
\mu &=& \frac{n}{1+\left(\alpha_1/6+\alpha_2/4+\alpha_3\right) R},
\label{musol}
\\
n^{-3/2}\chi_{(\varepsilon)}^{ab} &=& -2\alpha_1\left(R^{acbd}V_cV_d+\frac{R}{12}H^{ab}\right)
\nonumber \\ & & +\frac{\alpha_2}{2}
\left(R^{ab}-\frac{R}{4}g^{ab}\right),
\label{chiepssol}
\\
n^{3/2}\chi^{(\mu)}_{ab} &=& 2\alpha_1\left(R_{acbd}V^cV^d+\frac{R}{12}H_{ab}\right)
\nonumber \\ & & +\frac{(4\alpha_1+\alpha_2)}{2}
\left(R^{ab}-\frac{R}{4}g^{ab}\right),
\label{chimusol}
\end{eqnarray}
where $V^a = n^{3/4} v^a$ is the 4-velocity of the medium normalized according to the effective metric $g_{ab}$
 and $H_{ab}:=g_{ab}+V_a V_b$. 
 In Eqs.~(\ref{chiepssol}) and (\ref{chimusol}) indices are lowered and raised
by the effective metric and its inverse. Notice that, unless $\alpha_1 = \alpha_2 = 0$ --- which implies $\chi_{(\varepsilon)}^{ab}=0=\chi^{(\mu)}_{ab}$ ---,
 as a consequence of
  $\chi_{(\varepsilon)}^{ab}v_b=0=\chi^{(\mu)}_{ab}v^b$, only geometries associated with $g_{ab}$ which can 
  be put in the form 
  given by Eq.~(\ref{gdown}) {\it and} 
  satisfying
\begin{eqnarray}
R^a_{\;\;b} v^b = \frac{R}{4}v^a,
\label{constr}
\end{eqnarray}
for some timelike 4-vector $v^a$,  can be emulated by these anisotropic media --- with 
$v^a$ then set as the medium's 4-velocity. 
Using Einstein's equations to map this constraint to
the stress-energy-momentum
tensor $T^{ab}$ of the corresponding gravitational source, we have that 
\begin{eqnarray}
T^a_{\;\;b} v^b = \frac{T}{4}v^a,
\label{Tconstr}
\end{eqnarray}
where, again, the effective metric and its inverse are used to lower and raise indices (and $T := T^a_{\;\;a}$).
One can easily check that in case of perfect fluids --- characterized by a proper energy density $\rho$ and
(isotropic) pressure $p$ ---, Eq.~(\ref{Tconstr}) is only satisfied for $p = -\rho$; i.e., for a 
cosmological-constant-type
``fluid.'' However, if one allows for sources 
with anisotropic pressures ($p_1,p_2,p_3$), described by the
stress-energy-momentum tensor
\begin{eqnarray}
T^{ab} = \rho u^a u^b +\sum_{j = 1}^3 p_j {\bf e}^a_j {\bf e}^b_j
\label{Tanisotropic}
\end{eqnarray}
--- with $\{u^a,{\bf e}^a_1,{\bf e}^a_2,{\bf e}^a_3\}$ being a tetrad and $u^a$  timelike ---, then
\begin{eqnarray}
\rho+\frac{1}{3}\sum_{j = 1}^3 p_j = 0
\label{rhoconstr}
\end{eqnarray}
and
\begin{eqnarray}
(V_a{\bf e}_j^a)\left(\rho+p_j\right) = 0,\;\;\;\;j = 1,2,3.
\label{pjconstr}
\end{eqnarray}
In particular, if $V^a = u^a$, then Eq.~(\ref{rhoconstr}) is the only additional constraint to be enforced.

Returning attention to the background effective  geometry
and recalling that all the geometric tensors are 
obtained from $g_{ab}$ given in Eq.~(\ref{gdown}), we see that Eq.~(\ref{constr}) actually comprises
 a system of four
differential equations which $n$ and $v^a$ must satisfy. Electromagnetism 
with nonminimal coupling described by Eq.~(\ref{QED}) can only be simulated in these anisotropic media if the
background spacetime geometry
is associated
to solutions of this
system [via Eq.~(\ref{gdown})]. We shall treat a particular solution 
to these differential equations later.

%
%
%
%
%

\section{Plane-symmetric anisotropic medium at rest} 
\label{sec:ani}


In this section, we consider the simplest case of an anisotropic medium: a plane-symmetric medium at rest in the
inertial lab frame. The purpose of this section is not yet to establish an analogy with some interesting gravitational
system, but to present the analysis in a simple context. In Sec.~\ref{sec:spherical} we apply the analysis to a more
appealing scenario.

Let us consider a medium at rest in an inertial laboratory,
such that in inertial Cartesian coordinates $\{(t,x,y,z)\} = {\mathbb R}^3\times {\cal I}\subseteq{\mathbb R}^4$ 
we have $v^\mu = (1,0,0,0)$,  $\mu=\mu(z)$, $\varepsilon=\varepsilon(z)$,   
\begin{eqnarray}
\chi_{(\varepsilon)}^{\alpha\beta}=(\Delta^{(\varepsilon)}/3)\,(2\delta_{z}^{\alpha}\delta_{z}^{\beta}-\delta_{x}^{\alpha}\delta_{x}^{\beta}
-\delta_{y}^{\alpha}\delta_{y}^{\beta}),
\label{chimu}
\end{eqnarray}
and
\begin{eqnarray}
\chi^{(\mu)}_{\alpha\beta}=(\Delta^{(\mu)}/3)\,(2\delta^{z}_{\alpha}\delta^{z}_{\beta}-\delta^{x}_{\alpha}\delta^{x}_{\beta}
-\delta^{y}_{\alpha}\delta^{y}_{\beta}), 
\label{chimu}
\end{eqnarray}
with $\Delta^{(\varepsilon)}=\Delta^{(\varepsilon)}(z)$, $\Delta^{(\mu)}=\Delta^{(\mu)}(z)$, $z\in {\cal I}$. 
(${\cal I}$ is an open real interval.) This simply means  that
\begin{align}
&D^j = \varepsilon_\bot E_j,\;\;j=x,y\;, \label{Dxy} \\
&D^z = \varepsilon_\parallel E_z, \label{Dxy} \\
&H_{j}=\mu_\bot^{-1}B^{j},\;\;j=x,y\;, \label{Hxy}\\
&H_z=\mu_\parallel^{-1}B^z, \label{Hz}
\end{align}
where $\varepsilon_\parallel -\varepsilon_\bot\equiv\Delta^{(\varepsilon)}$, $(2\varepsilon_\bot+\varepsilon_\parallel)/3\equiv\varepsilon$,
$\mu^{-1}_\parallel -\mu^{-1}_\bot\equiv\Delta^{(\mu)}$, and $(2\mu^{-1}_\bot+\mu_\parallel^{-1})/3\equiv\mu^{-1}$. 

In these coordinates, $g_{\mu \nu} = \sqrt{n}\;{\rm diag}(-n^{-2},1,1,1)$.
For convenience, we shall work in the generalized Coulomb gauge~\cite{Glauber91} in which $A_\mu =( 0,{\bf A})$ and 
$\partial_\bot \cdot(\varepsilon_\bot {\bf A}_\bot)+\partial_z(\varepsilon_\parallel A_z) = 0$, where we have defined 
${\bf A}_\bot :=(A_x,A_y)$, $\partial_\bot := (\partial_x,\partial_y)$. In this gauge, the $t$ component of 
Eq.~(\ref{mx1cov}) is automatically satisfied, while the spatial components lead to
\begin{eqnarray}
&&
\left[-\frac{\varepsilon_\bot}{\mu_\bot}\partial_t^2+\frac{1}{\mu_\bot\mu_\parallel}\partial_\bot^2
+\left(\frac{1}{\mu_\bot}\partial_z\right)^2\right] {\bf A}_\bot
\nonumber \\
&&\;\;\;\;
=\frac{1}{\mu_\bot\mu_\parallel}\left[\mu_\parallel\partial_z\left( \mu_\bot^{-1}\partial_\bot\right)-\varepsilon_\bot^{-1}\partial_z\left(\varepsilon_\parallel\partial_\bot\right)\right]A_z,
\label{Eqxy}\\
&&
\left[-\frac{\mu_\bot}{\varepsilon_\bot}\partial_t^2+\frac{1}{\varepsilon_\bot\varepsilon_\parallel}\partial_\bot^2
+\left(\frac{1}{\varepsilon_\bot}\partial_z\right)^2\right](\varepsilon_\parallel A_z) = 0. 
\label{Eqz}
\end{eqnarray} 

First, let us consider solutions $\bf A$ such that $A_z = 0$, which describe
electric fields which are perpendicular to the $z$ direction --- {\it transverse electric modes},
${\bf A}^{\rm (TE)}$, for 
short~\cite{Claudia2009}.
In this case, our gauge condition ensures that there exists a scalar field $\psi$ such that 
$A^{\rm (TE)}_x=\partial_y\psi$ and $A^{\rm (TE)}_y=-\partial_x\psi$. 
Moreover, making use of the staticity and planar symmetry of the present scenario, we can write
$\psi = e^{-i(\omega t -{\bf k}_\bot\cdot {\bf x}_\bot)} f^{\rm (TE)}_{\omega { k}_\bot}(z)$, where ${\bf x}_\bot = (x,y,0)$,
${\bf k}_\bot = (k_x,k_y,0)$, $k_\bot =\|{\bf k}_\bot\|$, and $f^{\rm (TE)}_{\omega {k}_\bot}(z)$ satisfies
\begin{equation}
\left[-\frac{d^2}{d\zeta^2}+\left(
\frac{{k}_\bot^2}{\mu_\bot \mu_\parallel}-\frac{\varepsilon_\bot \omega^2}{\mu_\bot}\right)\right]
f^{\rm (TE)}_{\omega{k}_\bot}=0,
\label{slTE}
\end{equation}
with $\zeta$ being a spatial coordinate
such that $d\zeta = \mu_\bot dz$. The Eq.~(\ref{slTE}) must be supplemented by
 boundary conditions for $f^{\rm (TE)}_{\omega{k}_\bot}$.
 Imposing Eq.~(\ref{bcgen}) to these modes leads to
 \begin{eqnarray}
\left.\left[\overline{f^{\rm (TE)}_{\omega{ k}_\bot}}\frac{d}{d\zeta}f^{\rm (TE)}_{\omega'{ k}_\bot}
-f^{\rm (TE)}_{\omega'{ k}_\bot}\frac{d}{d\zeta}\overline{f^{\rm (TE)}_{\omega{ k}_\bot}}\right]\right|_{\dot{\cal I}} = 0,
\label{bcplaneTE}
\end{eqnarray}
where $[ ]|_{\dot{\cal I}}$ denotes the flux of the quantity in square brackets through $\dot{\cal I}$.
This condition restricts the possible
values of $\omega^2$. Let ${\cal E}_{k_\bot}^{\rm (TE)}$ 
be the ($k_\bot$-dependent) set of $\omega$ values for which Eqs.~(\ref{slTE}) and (\ref{bcplaneTE})
are satisfied for $f^{\rm (TE)}_{\omega{ k}_\bot}\nequiv 0$.
For $\omega,\omega'\in {\cal E}^{\rm (TE)}_{k_\bot +}:= {\cal E}_{k_\bot}^{\rm (TE)} \cap {\mathbb R}^\ast_+$,
we can
orthonormalize these modes according to
\begin{eqnarray}
\left(A^{\rm (TE)}_{\omega{\bf k}_\bot},A^{\rm (TE)}_{\omega'{\bf k}'_\bot}\right)
& =& -\left(\overline{A^{\rm (TE)}_{\omega{\bf k}_\bot}},\overline{A^{\rm (TE)}_{\omega'{\bf k}'_\bot}}\right)\nonumber \\
&=&
\delta_{\omega\omega'}\,\delta({\bf k}_\bot-{\bf k}_\bot '),
\label{normTE}
\\
\left(A^{\rm (TE)}_{\omega{\bf k}_\bot},\overline{A^{\rm (TE)}_{\omega'{\bf k}'_\bot}}\right) &=& 
0
\label{normTETEbar}
\end{eqnarray}
($\delta_{\omega\omega'}$ being the appropriate Dirac-delta distribution on ${\cal E}_{k_\bot}^{\rm (TE)} $), 
where  the sesquilinear form given in Eq.~(\ref{sesq}), applied to the current scenario, takes the form
\begin{eqnarray}
\left(A,A'\right)&:=&i \int_{\Sigma} d^3x\,\left[\varepsilon_\bot
\left(\bar{\bf A}_\bot\cdot \partial_t{\bf A}_\bot'-{\bf A}_\bot'\cdot \partial_t\bar{\bf A}_\bot\right)\right.
\nonumber \\
& &\;\;\;\;\;\;\;\;\;\;\;\;\;\;\;
 \left.+\varepsilon_\parallel \left(\bar{ A}_z \partial_t{ A}_z'-{ A}_z' \partial_t\bar{ A}_z\right)\right]
\label{sesqscenario}
\end{eqnarray}
We obtain (up to a global phase)
\begin{eqnarray}
\A^{\rm (TE)}_{\omega \k_{\bot}}=\frac{{\bf k}_\bot \times{\bf n}_z}{2 \pi k_\bot\sqrt{2\omega}}
e^{-i(\omega t-\k_{\bot}\cdot\x_{\bot})}f^{\rm (TE)}_{\omega k_{\bot}}(z),
\label{ATEfinal}
\end{eqnarray}
with
\begin{equation}
\int_{\cal I} dz \,\varepsilon_\bot(z) \,\overline{f^{\rm (TE)}_{\omega k_{\bot}}}(z)\,f^{\rm (TE)}_{\omega' k_{\bot}}(z)=\delta_{\omega\omega'}
\label{ffTE}
\end{equation}
and ${\bf n}_z := (0,0,1)$.

The second set of solutions of Eqs.~(\ref{Eqxy}) and (\ref{Eqz}), which describe
magnetic fields which are perpendicular to the $z$ direction --- 
{\it transverse magnetic modes},
${\bf A}^{\rm (TM)}$, for 
short~\cite{Claudia2009} ---,
is obtained by conveniently setting $A^{\rm (TM)}_z=\varepsilon_\parallel^{-1}\partial^2_{\bot}\phi$, where $\phi$
is an auxiliary function. Our gauge condition then leads to $A_j^{\rm (TM)} = -\varepsilon_\bot^{-1}\partial_j\partial_z\phi$,
$j=x,y$.
Using, again, staticity and planar symmetry, we find solutions of the form $\phi = 
e^{-i(\omega t-{\bf k}_\bot\cdot{\bf x}_\bot)} f^{\rm (TM)}_{\omega {k}_\bot}(z)$,
where $f^{\rm (TM)}_{\omega {k}_\bot}(z)$ satisfies
\begin{equation}
\left[-\frac{d^2}{d\xi^2}+
\left(
\frac{{k}_\bot^2}{\varepsilon_\bot \varepsilon_\parallel}-\frac{\mu_\bot \omega^2}{\varepsilon_\bot}\right)\right]
f^{\rm (TM)}_{\omega{k}_\bot}=0,
\label{slTM}
\end{equation}
with $\xi$ being a spatial coordinate
such that $d\xi = \varepsilon_\bot dz$. 
The boundary condition
imposed by Eq.~(\ref{bcgen}) now leads to
 \begin{eqnarray}
\left.\left[
\omega^2\overline{f^{\rm (TM)}_{\omega{ k}_\bot}}\frac{d}{d\xi}f^{\rm (TM)}_{\omega'{ k}_\bot}
-
\omega'^2
f^{\rm (TM)}_{\omega'{ k}_\bot}\frac{d}{d\xi}\overline{f^{\rm (TM)}_{\omega{ k}_\bot}}\right]\right|_{\dot{\cal I}} = 0.
\label{bcplaneTM}
\end{eqnarray}
Let 
${\cal E}_{k_\bot}^{\rm (TM)} $ be the
($k_\bot$-dependent) set of $\omega$ values for which Eqs.~(\ref{slTM}) and (\ref{bcplaneTM})
are satisfied for $f^{\rm (TM)}_{\omega{ k}_\bot}\nequiv 0$.
For $\omega,\omega'\in {\cal E}_{k_\bot +}^{\rm (TM)} :=  {\cal E}_{k_\bot}^{\rm (TM)} \cap {\mathbb R}^\ast_+$,
we can
normalize these modes according to
\begin{eqnarray}
\left(A^{\rm (TM)}_{\omega{\bf k}_\bot},A^{\rm (TM)}_{\omega'{\bf k}'_\bot}\right) &=&-\left(\overline{A^{\rm (TM)}_{\omega{\bf k}_\bot}},\overline{A^{\rm (TM)}_{\omega'{\bf k}'_\bot}}\right) 
\nonumber \\
&=&
\delta_{\omega\omega'}\,\delta({\bf k}_\bot-{\bf k}_\bot '),
\label{normTM}
\\
\left(A^{\rm (TM)}_{\omega{\bf k}_\bot},\overline{A^{\rm (TM)}_{\omega'{\bf k}'_\bot}}\right) &=&0
\end{eqnarray}
($\delta_{\omega\omega'}$ now being the appropriate Dirac-delta distribution on ${\cal E}_{k_\bot}^{\rm (TM)} $), 
obtaining (up to a global phase)
\begin{eqnarray}
{\bf A}^{\rm (TM)}_{\omega \k_{\bot}}=\frac{e^{-i(\omega t-\k_{\bot}\cdot\x_{\bot})}}{2\pi k_\bot\sqrt{2\omega^3}}
\left(\frac{ k_\bot^2}{\varepsilon_\parallel}{\bf n}_z+i\frac{ {\bf k}_\bot}{\varepsilon_\bot}\frac{d}{dz}\right)
f^{\rm (TM)}_{\omega k_{\bot}}(z),
\label{ATMfinal}
\end{eqnarray}
with
\begin{equation}
\int_{\cal I} dz \,\mu_\bot(z) \,\overline{f^{\rm (TM)}_{\omega k_{\bot}}}(z)\,f^{\rm (TM)}_{\omega' k_{\bot}}(z)=\delta_{\omega\omega'}.
\label{ffTM}
\end{equation}
Moreover,  modes ${\bf A}^{\rm (TM)}_{\omega \k_{\bot}}$ and $\overline{{\bf A}^{\rm (TM)}_{\omega \k_{\bot}}}$ are orthogonal to modes
${\bf A}^{\rm (TE)}_{\omega \k_{\bot}}$ and $\overline{{\bf A}^{\rm (TE)}_{\omega \k_{\bot}}}$.

The solutions expressed in Eqs.~(\ref{ATEfinal}) and (\ref{ATMfinal}), dubbed {\it positive-frequency} normal modes, play a central role in
the construction of the Fock (Hilbert) space of the quantized  theory,
as described at the end of the previous section.
With these solutions, the quantum-field operator $\hat{\bf A}$ is represented by
\begin{eqnarray}
{\hat{\bf A}} = \sum_{{\rm J}\in\{{\rm TE},{\rm TM}\}}\int_{{\mathbb R}^2}d^2{\bf k}_\bot
\int_{{\cal E}_{k_\bot +}^{({\rm J})}} d\omega
\left[{\hat a}^{({\rm J})}_{\omega{\bf k}_\bot} {\bf A}^{\rm (J)}_{\omega \k_{\bot}}+ {\rm H.c.}\right],
\label{hatA}
\end{eqnarray}
where ``H.c.''~stands for ``Hermitian conjugate'' of the preceding term and ${\hat a}^{({\rm J})}_{\omega{\bf k}_\bot}$ (respectively, ${\hat a}^{({\rm J})\dagger}_{\omega{\bf k}_\bot} $)
is the annihilation (resp., creation) operator associated with mode $ {\bf A}^{\rm (J)}_{\omega \k_{\bot}}$, satisfying the canonical commutation relations:
\begin{eqnarray}
\left[{\hat a}^{({\rm J})}_{\omega{\bf k}_\bot},{\hat a}^{({\rm J}')\dagger}_{\omega'{\bf k}'_\bot}\right] &=& \delta^{\rm JJ'}
\delta_{\omega\omega'}\,\delta({\bf k}_\bot-{\bf k}_\bot '),
\label{CCR1}
\\
\left[{\hat a}^{({\rm J})}_{\omega{\bf k}_\bot},{\hat a}^{({\rm J}')}_{\omega'{\bf k}'_\bot}\right] & =& 0.
\label{CCR2}
\end{eqnarray}

As an application of our quantization scheme one can use the above formulas to obtain, for instance, the Carniglia-Mandel quantization~\cite{Carniglia} in a straightforward way. 
The system in this case is composed by a dielectric-vacuum interface at $z=0$ and a non-magnetizable ($\mu_{\|}=\mu_{\bot}=1$) homogeneous isotropic non-dispersive 
dielectric ($\varepsilon_{\|}=\varepsilon_{\bot}=\varepsilon\equiv n^2$) filling the half-space $z<0$. These data enter Eqs.~\eqref{slTE} and \eqref{slTM}, thus describing the background in terms of effective potentials of one-dimensional Schr\"odinger-like problems.

\subsection{Instability analysis} 
\label{subsec:insta}

In the analysis presented above, it was implicitly assumed that all constitutive functions
 $\varepsilon_\bot$, $\varepsilon_\parallel$, $\mu_\bot$, and $\mu_\parallel$ are 
 positive functions of $z\in {\cal I}$. This condition
 ensures that the field modes presented in Eqs.~(\ref{ATEfinal}) and (\ref{ATMfinal}), together with their complex conjugates,
 constitute a complete set of (complexified) solutions of Maxwell equations in ${\mathbb R}^3\times {\cal I}$; in other words, the boundary-value problems
 defined by
Eqs.~(\ref{slTE},\ref{bcplaneTE})
 and Eqs.~(\ref{slTM},\ref{bcplaneTM}) 
 admit solutions only for (a subset of) $\omega^2 > 0$. This is easily seen by interpreting them 
 as null-eigenvalue
 problems for the linear operators defined in the square brackets of Eqs.~(\ref{slTE})
 and (\ref{slTM}). Experience with Schr\"odinger-like equations teaches us that
 these equations 
 have solutions provided the associated effective potentials (terms in parentheses)
become sufficiently negative in a given region --- which implies $\omega^2>0$ and, typically,  the larger the $k_\bot^2$,
the larger the $\omega^2$.

 Here, however, 
 we shall consider a more interesting situation. 
 It has been known for almost two decades that materials can be engineered so that
 some of their constitutive functions can  assume negative values~\cite{Smith00,Shelby01,Padilla06,Poddubny2013,Caligiuri2016}. These 
 exotic materials have been termed metamaterials.
 In this case, the effective potentials appearing in Eqs.~(\ref{slTE}) and (\ref{slTM}) may become
 sufficiently negative --- granting solutions to these boundary-value problems --- without demanding $\omega^2> 0$.
 For instance, if $\mu_\parallel <0$ (with $\mu_\bot, \varepsilon_\bot>0$), then the larger the value of $k_\bot$, the more negatively it contributes to the effective potential
 of Eq.~(\ref{slTE}),
 favoring the appearance of solutions with
 smaller (possibly negative) values of $\omega^2$. The same is true for Eq.~(\ref{slTM}) if $\varepsilon_\parallel <0$ and
 similar analysis can be done if any other constitutive function becomes negative.
 
 At this point, we must introduce an element of reality concerning the constitutive functions. We have been treating these quantities as
 given functions of $z$ alone --- neglecting dispersion effects, since we are, here, interested in gravity analogues. 
 However,  these material properties generally depend on characteristics of the
 electromagnetic field itself, particularly on its time variation (i.e., on $\omega$), in which case 
  Eqs.~\eqref{DE} and \eqref{HB} would be valid mode by mode, with the  constitutive tensors
  $\varepsilon^{ab}$ and $\mu_{ab}$ possibly being different for different modes. When translated 
  to spacetime-dependent quantities, Eqs.~(\ref{DE}) and (\ref{HB}) would be substituted by sums over the set of allowed field modes~\cite{RV2020}. 
Therefore, the precise key assumption about our metamaterial
 media is that some of their anisotropic constitutive functions $\varepsilon_\bot$, $\varepsilon_\parallel$, $\mu_\bot$,  $\mu_\parallel$
can become negative for some $\omega$ on the positive imaginary axis, 
$\omega^2<0$. Notwithstanding, the less restrictive condition ${\rm Im}(\omega)> 0$ would suffice for our purposes. However,
dealing with the case ${\rm Im}(\omega) {\rm Re}(\omega)\neq 0$ would involve quantization in active media, which we shall treat elsewhere~\cite{RV2020}. Moreover,
our focus here is to show that the electromagnetic field itself can exhibit interesting behavior without need to exchange energy with the medium (which occurs
in dispersive/active media). This justifies our focus on $\omega^2<0$ in what follows.
The possibility of having this type of material will be
discussed later.

 Let  $\omega^2 =- \Omega^2 $ (with $\Omega> 0$) be such value for which at least one of the  constitutive functions is negative
 for $z\in{\cal I}$. 
 Thus, both the effective potentials of Eqs.~(\ref{slTE}) and (\ref{slTM}) take the general form
 \begin{eqnarray}
 V_{\text{\it eff}} = C_1 k^2_\bot + C_2 \Omega^2,
 \label{Veff}
 \end{eqnarray}
 with $C_1$ and $C_2$ being functions of $z$. Two interesting  possibilities arise:
 \\
 
 \noindent
 $\bullet$ {(i) {$C_1<0$}:}
 In this case, the larger the value of $k_\bot$, the more negative the effective potential gets. Therefore, it is 
 quite reasonable to expect that, for a given size of the interval ${\cal I}$,
 one can always find ``large enough'' values of
 $k_\bot$ --- certainly satisfying $k_\bot^2 > C_2 \Omega^2/|C_1|$ ---
 such that the Schr\"odinger-like equation with effective potential $V_{\text{\it eff}}$ admits
 null-eingenvalue solutions. We shall refer to
 this situation as {\it large-$k_\bot$ instability};
 \\
 
 \noindent
 $\bullet$ (ii) {{$C_1>0$ and $C_2 <0$}}: 
 Under these conditions, the effective potential $V_{\text{\it eff}}$, as a function of $k_\bot$, is bounded from below:
 $V_{\text{\it eff}} \geq -|C_2| \Omega^2$. Therefore, 
 a Schr\"odinger-like equation with effective potential $V_{\text{\it eff}}$ only admits
 null-eigenvalue solutions provided
 $k_\bot$ is ``sufficiently small'' --- certainly satisfying $k_\bot^2<|C_2|\Omega^2/C_1$ --- and
 the size of the interval where $V_{\text{\it eff}}$ is negative is ``sufficiently large.'' 
  We shall refer to this situation as {\it minimum-width instability}.
 \\

 Let us call  $g^{({\rm J})}_{\Omega k_\bot}$ the null-eigenvalue solutions mentioned in either case above, 
 with ${\rm J}\in\{{\rm TE},{\rm TM}\}$ depending on whether it refers to Eq.~(\ref{slTE})
 or~(\ref{slTM}) with $\omega^2 = -\Omega^2$ (without loss of generality, $\Omega >0$).
These solutions are associated with unstable electromagnetic modes whose temporal
behavior is proportional to $e^{\pm \Omega t}$.
Although it might be tempting not to consider these ``runaway'' solutions, \cite{Raabe2007,Huttner1992}, they are essential, if they exist,  to expand an arbitrary initial field configuration satisfying the boundary-value problems set by Eqs.~(\ref{slTE},\ref{bcplaneTE}) 
and (\ref{slTM},\ref{bcplaneTM}); in other words,
the stationary modes alone do not
constitute a complete set of solutions of Maxwell's equations with the given boundary conditions. And even if, on the classical level, one might want to restrict
attention to initial field configurations which have no contribution coming from these unstable modes --- which is certainly unnatural, for causality forbids
the system to constrain its initial configuration based on its future behavior ---, inevitable quantum fluctuations of these modes would grow,  making them 
dominant some time e-foldings ($t\sim N\Omega^{-1}$, $N\gg 1$) after the proper material conditions having been engineered.
Therefore, these modes are as physical as the 
oscillatory ones. In fact, artificial inconsistencies have been reported in the literature, regarding field
quantization in active media
\cite{Raabe2007,Huttner1992}, which are completely cured when unstable modes are included
in the analysis~\cite{RV2020}.

It is interesting to note that depending on which constitutive function is negative, Eqs.~(\ref{slTE}) and (\ref{slTM}) may incur in different types of instabilities. For instance,
if $\mu_\bot <0$ for a given $\omega^2 = -\Omega^2<0$, with all other constitutive functions being positive, then Eq.~(\ref{slTE}) exhibits case-(i) instability, while Eq.~(\ref{slTM})
incur in case-(ii) instability. This means that unstable TE modes --- with some  $k_\bot> \sqrt{\mu_\parallel \varepsilon_\bot}\, \Omega$ --- 
would certainly be present, while unstable TM modes --- with some $k_\bot<\sqrt{|\mu_\bot| \varepsilon_\parallel}\, \Omega$ --- would only appear if the width of the material (size of the
interval ${\cal I}$) is larger than some critical value. We shall illustrate these facts in a simple example below. But first, let us analyze some features of these unstable modes.
In order not to rely on particular initial field configurations, let us focus on the inevitable quantum fluctuations of these modes.
 
\subsubsection{Unstable TE modes}

Repeating the procedure which led us from Eq.~(\ref{slTE}) to Eq.~(\ref{ATEfinal})  for the stable modes, 
unstable TE modes, $A^{(u{\rm TE})}_{\Omega{\bf k}_\bot}$, properly orthonormalized according to
\begin{eqnarray}
\left(A^{(u{\rm TE})}_{\Omega{\bf k}_\bot},A^{(u{\rm TE})}_{\Omega'{\bf k}'_\bot}\right)
& =& -\left(\overline{A^{(u{\rm TE})}_{\Omega{\bf k}_\bot}},\overline{A^{(u{\rm TE})}_{\Omega'{\bf k}'_\bot}}\right)\nonumber \\
&=&
\delta_{\Omega\Omega'}\,\delta({\bf k}_\bot-{\bf k}_\bot '),
\label{normuTE}
\\
\left(A^{(u{\rm TE})}_{\Omega{\bf k}_\bot},\overline{A^{(u{\rm TE})}_{\Omega'{\bf k}'_\bot}}\right) &=& 
0,
\label{normuTEuTE}
\end{eqnarray}
(and orthogonal to all other modes) read (up to a time translation)
\begin{eqnarray}
\A^{(u{\rm TE})}_{\Omega \k_{\bot}}&=&\frac{{\bf k}_\bot \times {\bf n}_z}{4\pi k_\bot \sqrt{\Omega \sin\kappa}}
e^{i\k_{\bot}\cdot\x_{\bot}} g^{\rm (TE)}_{\Omega k_{\bot}}(z)\nonumber \\
& & \times
\left(e^{\Omega t -i s_{\varepsilon}^{\bot}\kappa/2} + e^{-\Omega t +is_{\varepsilon}^{\bot}\kappa/2} \right),
\label{AuTEfinal}
\end{eqnarray}
with  $0<\kappa <\pi$, $g^{\rm (TE)}_{\Omega k_{\bot}}$ 
normalized according to
\begin{eqnarray}
\left|\int_{\cal I} dz \,\varepsilon_\bot(z) \,\overline{g^{\rm (TE)}_{\Omega k_{\bot}}}(z)\,g^{\rm (TE)}_{\Omega' k_{\bot}}(z)\right|& =&\delta_{\Omega\Omega'},
\label{guTE}
\end{eqnarray}
and $s_\varepsilon^\bot$ being the sign of the integral above. 
Calculating the electric ${\bf E}_{\Omega {\bf k}_\bot}^{(u{\rm TE})}$ and magnetic ${\bf B}_{\Omega {\bf k}_\bot}^{(u{\rm TE})}$
fields associated to these modes, we have:
\begin{eqnarray}
\E^{(u{\rm TE})}_{\Omega \k_{\bot}}&=&\frac{\sqrt{\Omega}}{4\pi {k}_{\bot}\sqrt{\sin\kappa}}
({\bf n}_z\times {\bf k}_\bot)\,
e^{i\k_{\bot}\cdot\x_{\bot}}g^{\rm (TE)}_{\Omega k_{\bot}}(z)\nonumber \\
& & \times
\left(e^{\Omega t -is_\varepsilon^\bot\kappa/2} - e^{-\Omega t +is_\varepsilon^\bot\kappa/2} \right),
\label{EuTE}\\
\B^{(u{\rm TE})}_{\Omega \k_{\bot}}&=&\frac{e^{i\k_{\bot}\cdot\x_{\bot}}}{4\pi k_\bot\sqrt{\Omega \sin\kappa}}
\left(-i k_\bot^2 {\bf n}_z+{\bf k}_\bot\frac{d}{dz}\right)g^{\rm (TE)}_{\Omega k_{\bot}}(z)
\nonumber \\
& & \times
\left(e^{\Omega t -is_\varepsilon^\bot\kappa/2} + e^{-\Omega t +is_\varepsilon^\bot\kappa/2} \right).
\label{BuTE}
\end{eqnarray}


\subsubsection{Unstable TM modes}

Now, turning to the TM modes, we repeat the procedure which led us from Eq.~(\ref{slTM}) to Eq.~(\ref{ATMfinal})  for the stable modes. 
Unstable TM modes, $A^{(u{\rm TM})}_{\Omega{\bf k}_\bot}$, properly orthonormalized according to
\begin{eqnarray}
\left(A^{(u{\rm TM})}_{\Omega{\bf k}_\bot},A^{(u{\rm TM})}_{\Omega'{\bf k}'_\bot}\right)
& =& -\left(\overline{A^{(u{\rm TM})}_{\Omega{\bf k}_\bot}},\overline{A^{(u{\rm TM})}_{\Omega'{\bf k}'_\bot}}\right)\nonumber \\
&=&
\delta_{\Omega\Omega'}\,\delta({\bf k}_\bot-{\bf k}_\bot '),
\label{normuTM}
\\
\left(A^{(u{\rm TM})}_{\Omega{\bf k}_\bot},\overline{A^{(u{\rm TM})}_{\Omega'{\bf k}'_\bot}}\right) &=& 
0,
\label{normuTMuTM}
\end{eqnarray}
(and orthogonal to all other modes) read (up to a time translation)
\begin{eqnarray}
\A^{(u{\rm TM})}_{\Omega \k_{\bot}}&=&\frac{e^{i\k_{\bot}\cdot\x_{\bot}}}{4\pi k_\bot\sqrt{\Omega^3\sin\kappa}}
\left(\frac{ k_\bot^2}{\varepsilon_\parallel}{\bf n}_z+i\frac{{\bf k}_\bot}{\varepsilon_\bot }\frac{d}{dz}\right)g^{\rm (TM)}_{\Omega k_{\bot}}(z)
\nonumber \\
& &\times
\left(e^{\Omega t +is_\mu^\bot\kappa/2} + e^{-\Omega t -is_\mu^\bot\kappa/2} \right),
\label{AuTMfinal}
\end{eqnarray}
where, again,  $0<\kappa<\pi$,  
$g^{\rm (TM)}_{\Omega k_{\bot}}$ is
normalized according to
\begin{eqnarray}
\left|\int_{\cal I} dz \,\mu_\bot(z) \,\overline{g^{\rm (TM)}_{\Omega k_{\bot}}}(z)\,g^{\rm (TM)}_{\Omega' k_{\bot}}(z)
\right|
& =&\delta_{\Omega\Omega'},
\label{guTM}
\end{eqnarray}
and $s_\mu^\bot$ is the sign of the integral above.
Calculating the electric ${\bf E}_{\Omega {\bf k}_\bot}^{(u{\rm TM})}$ and magnetic ${\bf B}_{\Omega {\bf k}_\bot}^{(u{\rm TM})}$
fields associated to these modes, we have:
\begin{eqnarray}
\E^{(u{\rm TM})}_{\Omega \k_{\bot}}&=&-\frac{e^{i\k_{\bot}\cdot\x_{\bot}}}{4\pi k_\bot\sqrt{\Omega\sin\kappa}}
\left(\frac{ k_\bot^2}{\varepsilon_\parallel}{\bf n}_z+i\frac{ {\bf k}_\bot}{\varepsilon_\bot}\frac{d}{dz}\right)g^{\rm (TM)}_{\Omega k_{\bot}}(z)
\nonumber \\
& &\times
\left(e^{\Omega t +is_\mu^\bot\kappa/2} - e^{-\Omega t -is_\mu^\bot\kappa/2} \right),
\label{EuTM}\\
\B^{(u{\rm TM})}_{\Omega \k_{\bot}}&=&\frac{i 
\mu_\bot\sqrt{\Omega}}{4\pi k_\bot\sqrt{
\sin\kappa}}
( {\bf n}_z\times{\bf k}_\bot)\, e^{i\k_{\bot}\cdot\x_{\bot}}g^{\rm (TM)}_{\Omega 
k_{\bot}}(z)
\nonumber \\
& & \times
\left(e^{\Omega t +is_\mu^\bot\kappa/2} + e^{-\Omega t -is_\mu^\bot\kappa/2} \right).
\label{BuTM}
\end{eqnarray}

\vskip 1.0cm

The modes given by Eqs.~(\ref{AuTEfinal}) and (\ref{AuTMfinal}), if present, must be added 
to the expansion of the field operator $\hat{\bf A}$ given in Eq.~(\ref{hatA}), along with 
their complex conjugates --- 
with corresponding
annihilation $\hat{a}^{(u{\rm J})}_{\Omega \k_\bot}$ and creation $\hat{a}^{(u{\rm J})\dagger}_{\Omega \k_\bot}$
operators, ${\rm J}\in \{{\rm TE},{\rm TM}\}$. The resulting operator expansion  can then be used
to calculate electromagnetic-field fluctuations and correlations. In the presence of unstable modes,
it is easy to see that the field's vacuum fluctuations  are eventually ($t\gg \Omega^{-1}$) dominated
 by these exponentially-growing
 modes. Obviously, this instability cannot persist indefinitely as 
 these wild  fluctuations will affect the medium's properties, supposedly leading the whole
 system to a final stable state. In some gravitational contexts, stabilization 
 occurs by decoherence of these growing vacuum fluctuations~\cite{LLMV}, giving rise to a nonzero classical field configuration ---
  a phenomenon called spontaneous scalarization (for spin-$0$)~\cite{PCBRS,DE1,DE2,B} or vectorization (for spin-$1$ fields)~\cite{Cardoso2019}.
  It is possible that something similar might occur in the analogous system. 
 We shall discuss this point further in Sec.~\ref{sec:stabilization}.

\subsection{Example}
\label{subsec:explane}

Let us consider a very simple system just to illustrate the results above in a concrete scenario:
a slab of width $L$ (in the region $-L/2<z<L/2$), 
made of a homogeneous material with, say, $\mu_\bot<0$ for a given
$\omega^2 = -\Omega^2$ ($\Omega >0$) and all other
constitutive functions positive. 
For concreteness sake, here we assume that this value $\omega^2 = -\Omega^2$ is isolated and that it is the most negative
value of $\omega^2$  for which $\mu_\bot<0$. This latter assumption is merely a matter of choice, while the former only affects the measure
on the set of quantum numbers 
${\bf k}_\bot$: $\int d^2 {\bf k}_\bot \to\int d\theta \sum_{{ k}_\bot} \!2 \pi { k}_\bot/L_\bot$,
$\delta(\k_\bot -\k_\bot') \to L_\bot \delta_{k_\bot k_\bot'} \delta(\theta-\theta')/(2\pi k_\bot)$,
  where $L_\bot$ is the legth scale
associated with the area of the ``infinite'' slab ($L_\bot\gg L$).

\begin{figure}[t]
\includegraphics[scale=0.25]{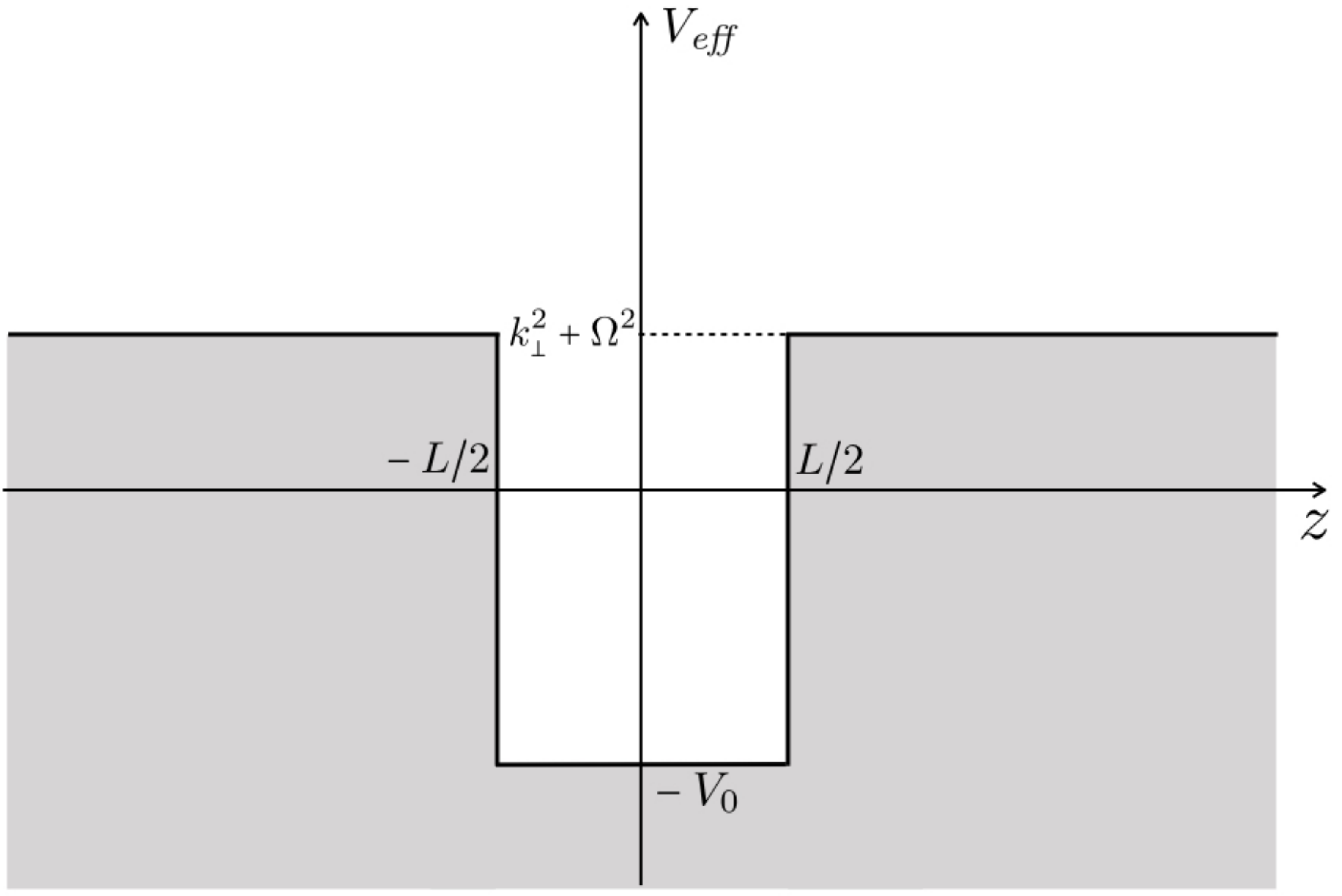}
\caption{Effective potential well which represents the homogeneous slab with negative $\mu_\bot$ for
the unstable electromagnetic modes.}
\label{fig:Veff}
\end{figure}

According to the discussion presented earlier, in this scenario,
 TE modes incur in case-(i)
(large-$k_\bot$) instability, while TM modes  undergo case-(ii) (minimum-width) instability. 
The solutions $g^{\rm (J)}_{\Omega k_\bot}$ of Eqs.~(\ref{slTE}) and (\ref{slTM}) with $\omega^2 = -\Omega^2$
are given by the normalizable ---
according to Eqs.~(\ref{guTE}) and (\ref{guTM}) --- solutions of the null-eigenvalue, Schr\"odinger-like 
equation
$$
\left(-\frac{d^2}{dz^2}+ V_{\it eff}\right)g^{\rm (J)}_{\Omega k_\bot} = 0,
$$
with $V_{\it eff}$ being the well potential represented in Fig.~\ref{fig:Veff}.
The depth of the potential is given by
\begin{eqnarray}
V_0 = \left\{
\begin{array}{ll}
|\mu_\bot| \varepsilon_\bot \Omega^2+\frac{|\mu_\bot|}{\mu_\parallel} k_\bot^2&,\;\; {\rm J} = {\rm TE}\\
|\mu_\bot| \varepsilon_\bot \Omega^2-\frac{\varepsilon_\bot}{\varepsilon_\parallel} k_\bot^2&, \;\;{\rm J} 
= {\rm TM}
\end{array}
\right. .
\label{eq:Veff}
\end{eqnarray}
Although here we focus only on unstable modes, associated with $g^{\rm (J)}_{\Omega k_\bot}$, note that in this example there would  also appear
stationary bound solutions associated with $f^{\rm (TE)}_{\omega_0 k_\bot}$ --- if $\mu_\bot < 0$ for some $\omega_0\in {\mathbb R}$ ---, for some
$k_\bot^2 > \max\{\omega_0^2,(n_\bot^{\rm (TE)}\omega_0)^2\}$, where $n_\bot^{\rm (TE)} := \sqrt{\mu_\parallel\varepsilon_\bot}$ is the transverse
refractive index for the TE modes. For such a hypothetical mode, the slab would act as a waveguide, keeping the mode confined due to total internal
reflections at its boundaries. The only peculiar feature here is that $k_\bot$ would assume arbitrarily large values (in practice, limited only by the
inverse length scale below which the continuous-medium idealization breaks down) for a given $\omega_0$.

Back to the unstable modes, a straightforward 
calculation leads to the
familiar even and odd solutions to the square-well potential, with $g^{\rm (J)}_{\Omega k_\bot}(z)$ exponentially
supressed for $|z|>L/2$ and
\begin{eqnarray}
g^{\rm (J)}_{\Omega k_\bot}(z) = \left\{
\begin{array}{ll}
\frac{{\cal N}^{\rm (J)}_{m}}{\cos a_{m}}\cos(2a_{m} z/L)&,\;0\leq m\;\;\text{even}\\ 
\frac{{\cal N}^{\rm (J)}_{m}}{\sin a_{m}}\sin(2a_{m} z/L) &,\;1\leq m\;\;\text{odd}
\end{array}
\right. 
\label{gsolex1}
\end{eqnarray}
($-L/2\leq z\leq L/2$),
where ${\cal N}^{\rm (J)}_m$ are normalization constants and, for the TE modes, $a_m\geq \Omega L |n_\parallel|/2$ are solutions of the transcendental equations
\begin{eqnarray}
 \sqrt{|\mu_\bot|\mu_\parallel}\sqrt{1-\frac{\Omega^2 L^2}{4a_m^2}|n_\parallel^2 |[1-(n_\bot^{\rm (TE)})^{-2}]} \nonumber
 \\
 =\left\{
 \begin{array}{ll}
 -\tan a_m&,\;m\text{ even}\\ 
 \cot a_m&,\;m\text{ odd}
 \end{array}
 \right. ,
\label{amTE}
\end{eqnarray}
while for the TM modes, $0\leq a_m\leq \Omega L |n_\parallel|/2$ and 
\begin{eqnarray}
 \sqrt{\varepsilon_\bot\varepsilon_\parallel}\sqrt{\frac{\Omega^2 L^2}{4a_m^2}|n_\parallel^2 |[1+|n_\bot^{\rm (TM)}|^{-2}]-1} \nonumber
 \\
 =\left\{
 \begin{array}{ll}
 \tan a_m&,\;m\text{ even}\\
 -\cot a_m&,\;m\text{ odd}
 \end{array}
 \right. .
\label{amTM}
\end{eqnarray}
The transverse momentum $k_\bot$ is given in terms of $a_m$ by
\begin{eqnarray}
& & k_\bot=k_\bot^{(m)}:=\left\{
\begin{array}{ll}
\frac{2}{L}\sqrt{\frac{\mu_\parallel}{|\mu_\bot|}\left(a_m^2-|n_\parallel^2|\frac{\Omega^2 L^2}{4}\right)} &,\;\text{TE modes}
\\
\frac{2}{L}\sqrt{\frac{\varepsilon_\parallel}{\varepsilon_\bot}\left(|n_\parallel^2|\frac{\Omega^2 L^2}{4}-a_m^2\right)} &,\;\text{TM modes}
\end{array}\right. .
\label{kbotnumsol}
\end{eqnarray}

\begin{figure}[t]
\includegraphics[scale=0.28]{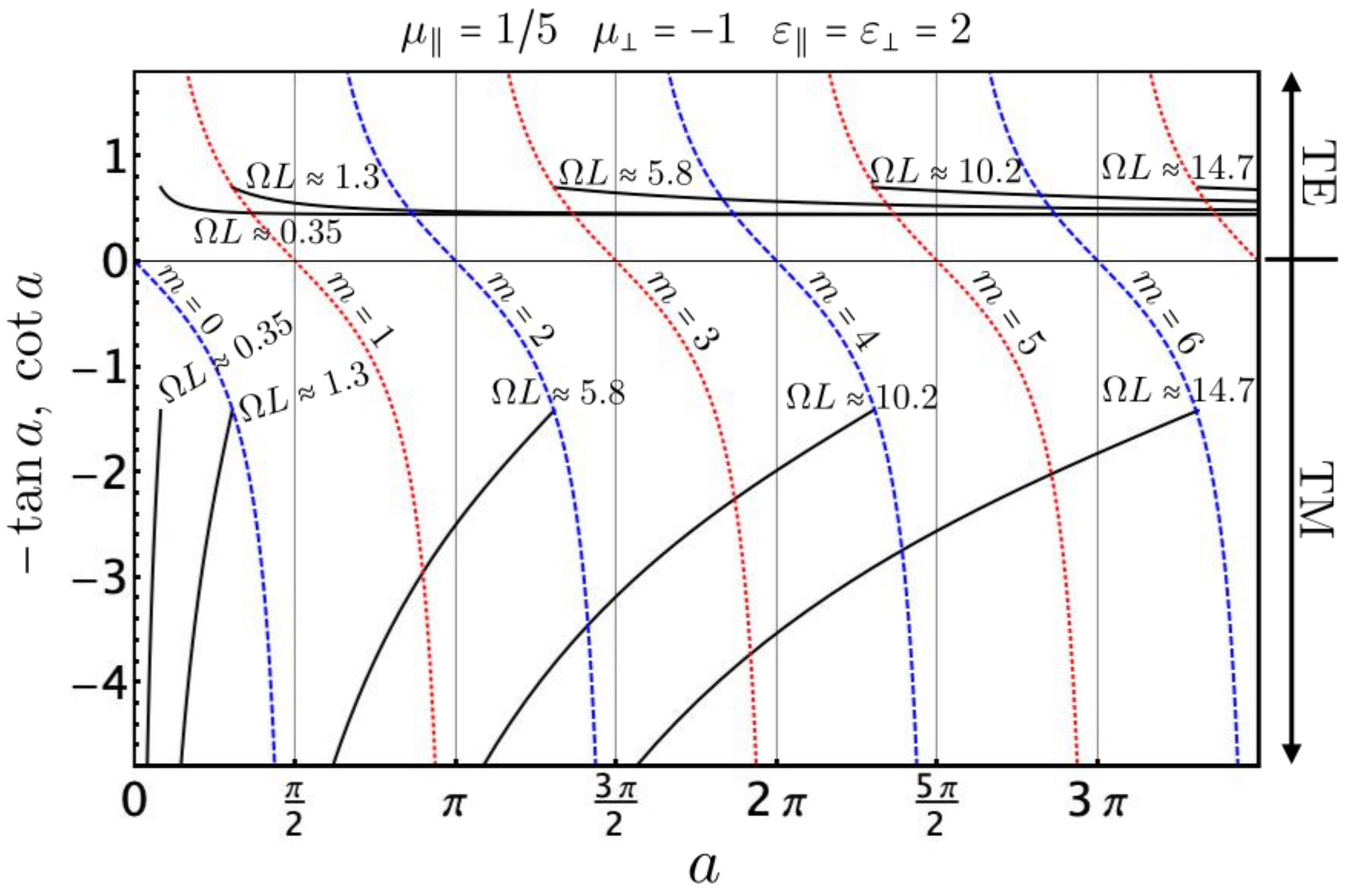}
\caption{Graphic representation of solutions of Eqs.~(\ref{amTE}) and (\ref{amTM}). The solid black curves
in the upper (respectively, lower) half plane represent the left-hand side (l.h.s.)~of Eq.~(\ref{amTE}) [resp., minus
the l.h.s.\ of Eq.~(\ref{amTM})] --- with $a_m$ replaced by the variable $a$ ---, for different values of $\Omega L$.
The dashed blue lines
(resp., dotted red lines) represent the function $-\tan a$ (resp., $\cot a$). The values $a_m$
appearing in Eq.~(\ref{gsolex1}) are determined by the crossing of the corresponding solid black curve with 
the dashed blue lines (for $m$ even) and the dotted red lines (for $m$ odd).}
\label{fig:numsol}
\end{figure}

\noindent
The explicit form of ${\cal N}_m^{\rm (J)}$ is not particularly important,
so we only present  its asymptotic behavior for
$k_\bot \to \infty$ for the TE modes,
\begin{eqnarray}
{\cal N}_m^{\rm (TE)} \approx 
\left\{
\begin{array}{ll}
\sqrt{\frac{2(1+|\mu_\bot |\mu_\parallel)}{L \varepsilon_\bot}}, & m \text{ even}\\
\sqrt{\frac{2(1+|\mu_\bot |\mu_\parallel)}{L \varepsilon_\bot |\mu_\bot |\mu_\parallel}}, & m \text{ odd}
\end{array}
\right., \; k_\bot \gg \Omega,
\label{NTEinfty}
\end{eqnarray}
and for $k_\bot \to 0$ for both TE and TM modes,
\begin{eqnarray}
{\cal N}_m^{\rm (TE)} \approx 
\left\{
\begin{array}{ll}
\sqrt{\frac{2(\varepsilon_\bot+|\mu_\bot |)}{L \varepsilon_\bot^2}}, & m \text{ even}\\
\sqrt{\frac{2(\varepsilon_\bot+|\mu_\bot |)}{L \varepsilon_\bot |\mu_\bot |}}, & m \text{ odd}
\end{array}
\right., \; k_\bot \ll \Omega,
\label{NTE0}
\\
{\cal N}_m^{\rm (TM)} \approx 
\left\{
\begin{array}{ll}
\sqrt{\frac{2(\varepsilon_\bot+|\mu_\bot |)}{L |\mu_\bot|^2}}, & m \text{ even}\\
\sqrt{\frac{2(\varepsilon_\bot+|\mu_\bot |)}{L \varepsilon_\bot |\mu_\bot |}}, & m \text{ odd}
\end{array}
\right., \; k_\bot \ll \Omega.
\label{NTM0}
\end{eqnarray}

In Fig.~\ref{fig:numsol}, we plot --- for different values of $\Omega L$ and given values of $\mu_\parallel$, $\mu_\bot$, $\varepsilon_\parallel$, and
$\varepsilon_\bot$ ---  the left-hand side of Eq.~(\ref{amTE}) (solid black curves in the upper half plane),
minus the left-hand side of Eq.~(\ref{amTM}) (solid black curves in the lower half plane) --- substituting, in both, $a_m$ by the variable $a$ ---, and
the functions $-\tan a$ and $\cot a$ (blue dashed lines and red dotted lines, respectively). Crossing of the blue dashed lines (respectively, red dotted lines) with
 a fixed solid black curve determines values $a = a_m$ for   even (resp., odd) solutions $g_{\Omega k_\bot}^{\rm (J)}$, for the corresponding value of $\Omega L$.
 The figure clearly corroborates our preliminary analysis, showing that unstable TE modes appear with arbitrarily large values of $a_m$ (and, therefore, of $k_\bot$)
 and that unstable TM modes only appear if $L$ is larger than some minimum width $L_0$, given by
 \begin{eqnarray}
 L_{0} = \frac{2\Omega^{-1}}{|n_\parallel |}\tan^{-1}\left(\sqrt{\frac{\varepsilon_\bot}{|\mu_\bot|}}\right).
 \label{lmin}
 \end{eqnarray}
 The unstable TE and TM modes inside the slab  can then be put in the form
 \begin{widetext}
 \begin{eqnarray}
\A^{(u{\rm TE})}_{\Omega \k^{(m)}_{\bot}}&=&\frac{{\cal N}^{\rm (TE)}_m({\bf k}^{(m)}_\bot \times {\bf n}_z)}{4
 \pi k_\bot^{(m)} \sqrt{\Omega \sin\kappa_m}} \frac{\cos\left(2a_m z/L+m\pi/2\right)}{\cos\left(a_m+m\pi/2\right)}
 e^{i\k^{(m)}_{\bot}\cdot\x_{\bot}}
\left(e^{\Omega t -i \kappa_m/2} + e^{-\Omega t +i\kappa_m/2} \right),\;\;m\geq m^{\rm (TE)},
\label{uTEex1}
\\
{\bf A}^{(u{\rm TM})}_{\Omega{\bf k}^{(m)}_\bot}
&=&
\frac{{\cal N}^{\rm (TM)}_m e^{i\k_{\bot}^{(m)}\cdot\x_{\bot}}}{4 \pi   \sqrt{\Omega^3\sin\kappa_m}}
\left(e^{\Omega t -i\kappa_m/2} + e^{-\Omega t+i\kappa_m/2} \right)\nonumber \\
& &\times
\left[
\frac{k_\bot^{(m)}{\bf n}_z}{\varepsilon_\parallel}\frac{\cos\left(2a_m z/L+m\pi/2\right)}{\cos\left(a_m+m\pi/2\right)}
-i\frac{{\bf k}^{(m)}_\bot}{\varepsilon_\bot k^{(m)}_\bot}\frac{2a_m}{L}
\frac{\sin\left(2a_mz/L+m\pi/2\right)}{\cos\left(a_m+m\pi/2\right)}
\right],\;\;
0\leq m\leq m^{\rm (TM)},
\label{uTMex1}
\end{eqnarray}
\end{widetext}
with
\begin{eqnarray}
m^{\rm (TE)} &:=&\left\lceil 1+\left(\frac{L}{L_0}-1\right)\frac{2}{\pi}\tan^{-1}
\left(\sqrt{\frac{\varepsilon_\bot}{|\mu_\bot |}}\right)\right\rceil,
\label{mTE}\\
m^{\rm (TM)}& :=&\left\lfloor\left(\frac{L}{L_0}-1\right)\frac{2}{\pi}\tan^{-1}
\left(\sqrt{\frac{\varepsilon_\bot}{|\mu_\bot |}}\right)\right\rfloor
\label{mTM}
\end{eqnarray}
($\lceil x \rceil$ represents the smallest integer 
larger than, or equal to, $x$, while
$\lfloor x\rfloor$ represents the largest integer smaller than, or
equal to, $x$).
The corresponding electric and magnetic field modes are
\begin{widetext}
\begin{eqnarray}
{\bf E}^{(u{\rm TE})}_{\Omega \k^{(m)}_{\bot}}&=&\frac{{\cal N}^{\rm (TE)}_m( {\bf n}_z \times {\bf k}^{(m)}_\bot)\sqrt{\Omega}}{4
 \pi k_\bot^{(m)} \sqrt{ \sin\kappa_m}}\frac{\cos\left(2a_m z/L+m\pi/2\right)}{\cos\left(a_m+m\pi/2\right)}
 e^{i\k^{(m)}_{\bot}\cdot\x_{\bot}}
\left(e^{\Omega t -i \kappa_m/2} - e^{-\Omega t +i\kappa_m/2} \right),\;\;m\geq m^{\rm (TE)},
\label{EuTEex1}
\\
{\bf B}^{(u{\rm TE})}_{\Omega{\bf k}^{(m)}_\bot}
&=&
\frac{{\cal N}^{\rm (TE)}_m e^{i\k_{\bot}\cdot\x_{\bot}}}{4 \pi  \sqrt{\Omega\sin\kappa_m}}
\left(e^{\Omega t -i\kappa_m/2} + e^{-\Omega t+i\kappa_m/2} \right)\nonumber \\
& &\times
\left[
-i k^{(m)}_\bot{\bf n}_z\frac{\cos\left(2a_m z/L+m\pi/2\right)}{\cos\left(a_m+m\pi/2\right)}
-\frac{{\bf k}^{(m)}_\bot}{ k^{(m)}_\bot}\frac{2a_m}{L}
\frac{\sin\left(2a_mz/L+m\pi/2\right)}{\cos\left(a_m+m\pi/2\right)}
\right],\;\;m\geq m^{\rm (TE)},
\label{BuTEex1}
\\
{\bf E}^{(u{\rm TM})}_{\Omega{\bf k}^{(m)}_\bot}
&=&
\frac{-{\cal N}^{\rm (TM)}_m e^{i\k^{(m)}_{\bot}\cdot\x_{\bot}}}{4\pi  \sqrt{\Omega\sin\kappa_m}}
\left(e^{\Omega t -i\kappa_m/2} - e^{-\Omega t+i\kappa_m/2} \right)\nonumber \\
& &\times
\left[
\frac{k^{(m)}_\bot{\bf n}_z}{\varepsilon_\parallel}\frac{\cos\left(2a_m z/L+m\pi/2\right)}{\cos\left(a_m+m\pi/2\right)}
-i\frac{{\bf k}^{(m)}_\bot}{\varepsilon_\bot k^{(m)}_\bot}\frac{2a_m}{L}
\frac{\sin\left(2a_mz/L+m\pi/2\right)}{\cos\left(a_m+m\pi/2\right)}
\right],\;\;0\leq m\leq m^{\rm (TM)},
\label{EuTMex1N}
\\
{\bf B}^{(u{\rm TM})}_{\Omega \k_{\bot}^{(m)}}&=&\frac{i {\cal N}^{\rm (TM)}_m\mu_\bot\sqrt{\Omega}
( {\bf n}_z\times {\bf k}^{(m)}_\bot )}{4
 \pi k_\bot^{(m)}\sqrt{\sin\kappa_m}} 
  e^{i\k_{\bot}^{(m)}\cdot\x_{\bot}}
 \frac{\cos\left(2a_m z/L+m\pi/2\right)}{\cos\left(a_m+m\pi/2\right)}
\left(e^{\Omega t -i \kappa_m/2} + e^{-\Omega t +i\kappa_m/2} \right),\;\;0\leq m\leq m^{\rm (TM)}.
\label{BuTMex1}
\end{eqnarray}
\end{widetext}

Let us recall that these modes give information about fluctuations and correlations of the electromagnetic field;
as long as decoherence does not come into play, the expectation values of the field are null, $\langle {\hat {\bf A}}\rangle = 
\langle {\hat {\bf E}}\rangle=\langle {\hat {\bf B}}\rangle={\bf 0}$.
We shall use these  modes later, when discussing  possible consequences
of these analogue instabilities. But first, let us explore more interesting analogies.

\section{Spherically-symmetric, stationary anisotropic  medium}
\label{sec:spherical}

In the previous section, we presented with great amount of detail the canonical quantization scheme for the electromagnetic field in flat spacetime in the presence of arbitrary plane-symmetric anisotropic
 polarizable/magnetizable media at linear order. The vacuum of such system was then identified with the vacuum of some nonminimally-coupled spin-$1$ field in a true curved spacetime described by the 
effective metric $g_{\alpha\beta}=\sqrt{n}\ \mbox{diag}(-n^{-2},1,1,1)$. The  analysis had the advantage of generalizing in a unified language the quantization of various interesting models 
coming from quantum optics in terms of simple equations (e.g., the Carniglia-Mandel modes~\cite{Carniglia}). 
However, the analogue spacetime for these configurations is of mathematical interest only and does not capture the symmetry of physical spacetimes. In order to study more appealing analogues, 
in this section we turn to spherically symmetric configurations, presenting them in a more concise way --- for the nuances of the quantization were already explained previously. 
In this context, we may obtain interesting analogues by also assuming that the medium is able to flow. 
If the refractive index in a flowing material is high enough, such that the velocity of light becomes smaller than the medium's velocity, then it is clear that a sort of event horizon will form 
(restricted only to some frequency band which may contain unstable modes). This kind of phenomenon enable us to study analogues of unstable black holes, for instance.

We start working in standard spherical coordinates $(t,r,\theta,\varphi)$, such that $\eta_{\mu\nu}=\mbox{diag}(-1,1,r^2,r^2\sin^2\theta)$. Let the medium's four-velocity field be $v^{\mu}=\gamma(1,v,0,0)$, 
where $v=v(r)$ and $\gamma=(1-v^2)^{-1/2}$. The effective-metric components then take the form
\begin{eqnarray}
g_{\alpha \beta} =\sqrt{n} \left(
\begin{array}{cccc}
-\gamma^2 (n^{-2}- v^2)& -(1-n^{-2})\gamma^2 v & 0 & 0\\
 -(1-n^{-2})\gamma^2 v & \gamma^2 (1-n^{-2} v^2) & 0 & 0\\
 0 & 0& r^2 & 0\\
 0& 0& 0& r^2 \sin^2\theta
\end{array}
\right),\nonumber \\
\label{gmunuss}
\end{eqnarray}
where the isotropic parts of the constitutive tensors 
(in the local, instantaneous rest frame of the medium) are functions of $r$ ---
$\varepsilon=\varepsilon(r)$, $\mu=\mu(r)$ --- and, as usual, $n^2 = \mu \varepsilon$. 
As for the traceless anisotropic tensors $\chi_{(\varepsilon)}^{ab}$ and $\chi^{(\mu)}_{ab}$, their
components read
\begin{eqnarray}
\chi_{(\varepsilon)}^{\alpha\beta}=(\Delta^{(\varepsilon)}/3)&(2\gamma^2v^2\delta_{t}^{\alpha}\delta_{t}^{\beta}+4\gamma^2v\delta_{t}^{(\alpha}\delta_{r}^{\beta)}+2\gamma^2\delta_{r}^{\alpha}\delta_{r}^{\beta}\nonumber\\
&-\delta^{\alpha}_{\theta}\delta^{\beta}_{\theta}r^{-2}-\delta^{\alpha}_{\varphi}\delta^{\beta}_{\varphi}r^{-2}\sin^{-2}\theta)
\label{sec4eq1}
\end{eqnarray}
and
\begin{eqnarray}
\chi^{(\mu)}_{\alpha\beta}=(\Delta^{(\mu)}/3)&(2\gamma^2v^2\delta^{t}_{\alpha}\delta^{t}_{\beta}-4\gamma^2v\delta^{t}_{(\alpha}\delta^{r}_{\beta)}+2\gamma^2\delta^{r}_{\alpha}\delta^{r}_{\beta}\nonumber\\
&-\delta_{\alpha}^{\theta}\delta_{\beta}^{\theta}r^2-\delta_{\alpha}^{\varphi}\delta_{\beta}^{\varphi}r^2\sin^2\theta).
\label{sec4eq2}
\end{eqnarray}
Similarly to the plane-symmetric case, these anisotropic tensors simply mean that in the instantaneous
local rest frame of the medium, its
electric permitivity and
magnetic permeability in the radial direction ($\varepsilon_\parallel$ and $\mu_\parallel$) and in the
angular directions ($\varepsilon_\bot$ and $\mu_\bot$) satisfy the same relations given below
Eqs.~(\ref{Dxy}-\ref{Hz}): $\varepsilon_\parallel -\varepsilon_\bot\equiv\Delta^{(\varepsilon)}$, $(2\varepsilon_\bot+\varepsilon_\parallel)/3\equiv\varepsilon$,
$\mu^{-1}_\parallel -\mu^{-1}_\bot\equiv\Delta^{(\mu)}$, and $(2\mu^{-1}_\bot+\mu_\parallel^{-1})/3\equiv\mu^{-1}$.



Not surprisingly, the lab coordinates $(t,r,\theta,\varphi)$ are not the most convenient ones to express 
Eqs.~(\ref{mx2cov}) and (\ref{mx1cov}) in the case of 
a moving medium. One might initially think  that coordinates $(\tau,{r},\theta,\varphi)$
which diagonalize the 
components of the
effective metric, obtained by defining
$\tau:=t-p(r)$, with $p(r)$ 
satisfying
\begin{equation}
\frac{dp}{dr}= -\frac{(n^2-1)v}{1-n^2v^2},\label{sec4eq5}
\end{equation} 
would lead to the simplest form of the field equations. In these coordinates, 
the effective line element $ds^2_{\text{\it eff}}$ becomes
\begin{equation}
ds^2_{\text{\it eff}}=\sqrt{n}\left[-n^{-2}Fd\tau^2+F^{-1}dr^2+
r^2(d\theta^2+\sin^2\theta d\varphi^2)\right],\label{sec4eq6}
\end{equation}
where $F=\gamma^2(1-n^2v^2)$. It is noteworthy that for 
$n = {\rm constant} >0$ (such that the factors of $n$ in 
$ds^2_{\text{\it eff}}$ 
can be absorbed via $\tau\mapsto n^{3/4}\tau$ and $r\mapsto n^{-1/4} r$), then the line element above can be made to
represent Schwarzschild spacetime by tuning $v$ so that $F \equiv (1-r_s/r)$,
where $r_s$ is some positive constant. This is achieved by a velocity field satisfying
 $v^2 =\left[1+(n^2-1) r/r_s\right]^{-1} $ ($n\neq 1$).

Despite this apparent simplification, the coordinate $\tau = t-p(r)$ 
with $p$ satisfying Eq.~(\ref{sec4eq5}) 
is not
convenient to express Maxwell's equations in anisotropic media. This is due to the kinematic polarization 
(resp., magnetization) caused by the magnetic (resp., electric) field. In the case of small velocities 
and isotropic materials, this effect is modeled by Minkowski's
 equations \cite{Minkowski1910}. The coordinates $(\tau,{r},\theta,\varphi)$ defined using
Eq.~\eqref{sec4eq5} ``diagonalizes'' only the isotropic part of the theory and do not take into account the 
anisotropies.
It turns out that  a much better choice is obtained by 
setting $\tau := t - p(r)$ and  replacing condition given in Eq.~(\ref{sec4eq5}) by
\begin{eqnarray}
\frac{dp}{dr}& =& -\frac{(n_\parallel^2-1)v}{1-n_\parallel^2 v^2},\label{sec4eq7}
\label{dpdr}
\end{eqnarray}
where, again, $n_\parallel^2: = \mu_\bot \varepsilon_\bot$. This choice fully decouples the electromagnetic
field modes in  the anisotropic, moving material medium, as we shall see below.

%
%

Introducing again the 4-potential $A_\mu$ via $F_{\mu\nu} = \partial_\mu A_\nu -\partial_\nu A_\mu$,
in these new coordinates $(\tau,r,\theta,\varphi)$, the convenient (generalized Coulomb) gauge
conditions read $A_\tau = 0$ and
\begin{eqnarray}
\partial_{\varrho}\left(\varepsilon_\parallel r^2 
A_{r}\right)+
\partial_\bot\cdot {\bf A}_\bot = 0,
\label{gCgcss}
\end{eqnarray}
where $\varrho$ is merely an auxiliary variable such that $dr/d\varrho \equiv\gamma^2
(1-n_\parallel^2v^2)/\varepsilon_\bot$,
 ${\bf A}_\bot = (A_\theta,A_\varphi)$, $\partial_\bot$ is the 
derivative operator on the unit sphere compatible with its metric, and it is understood that 
 $r$ is a function of the auxiliary variable $\varrho$. In this gauge, Maxwell's equations lead to 
\begin{eqnarray}
& & \left[-\frac{\mu_\bot}{\varepsilon_\bot}\partial_\tau^2 + \partial_\varrho^2+\frac{\gamma^2(1-n_\parallel^2v^2)}{\varepsilon_\bot\varepsilon_\parallel r^2}\Delta_S^{\!(0)}\right]\left(\varepsilon_\parallel r^2 
A_{r}\right) = 0,\;\;
\label{MEr}
\\
& & \left[-\frac{\varepsilon_\bot}{\mu_\bot}\partial_\tau^2 +
 \partial_\rho^2+\frac{\gamma^2(1-n_\parallel^2v^2)}{\mu_\bot\mu_\parallel r^2}\left(\Delta_S^{\!(1)}-1\right)\right] 
 {\bf A}_\bot 
 \nonumber \\
 & & = \partial_\bot\left[\partial_\rho\left(\frac{dr}{d\rho}A_r
\right)-\frac{\mu_\bot}{r^2\mu_\parallel \varepsilon_\bot}\frac{dr}{d\rho}\partial_\rho\left(\varepsilon_\parallel r^2A_r\right)
\right],
\label{MEtheta}
\end{eqnarray}
where $\rho$ appearing in Eq.~(\ref{MEtheta}) is another auxiliary variable defined through $dr/d\rho 
 \equiv\gamma^2(1-n_\parallel^2v^2)/\mu_\bot$ and $\Delta_S^{(0)}$ and $\Delta_S^{(1)}$ are the 
 Laplacian operators defined on the unit sphere, acting on scalar and covector fields,
 respectively. 
 
In order to solve these equations, we proceed in close analogy to the plane-symmetric case.
First, let us find solutions with $A_r = 0$ --- the transverse electric modes, ${\bf A}^{\rm (TE)} $.
The gauge conditions imply that these solutions can be written as
 ${\bf A}^{\rm (TE)} = \left(0, \partial_\varphi \psi/\sin\theta, -\sin\theta\partial_\theta \psi\right)$, where
 $\psi$ is an auxiliary function to be determined. Making use of the stationarity and spherical
 symmetry of the present scenario, we can look for field modes of the form
 $\psi= e^{-i\omega \tau} Y_{\ell m}(\theta,\varphi) f_{\omega \ell }^{\rm (TE)}(r)$, where
 $Y_{\ell m}$ are the scalar spherical harmonics. Substituting this into Eq.~(\ref{MEtheta}),
 $f^{\rm (TE)}_{\omega \ell}$ must satisfy
 \begin{eqnarray}
 \left[-\frac{d^2}{d\rho^2}+\left(\frac{\gamma^2(1-n_\parallel^2v^2)\ell(\ell+1)}
 {r^2 \mu_\bot \mu_\parallel}-\frac{\varepsilon_\bot \omega^2}{\mu_\bot}\right)\right] 
 f^{\rm (TE)}_{\omega \ell} = 0,
 \label{fTEss}
 \end{eqnarray}
where it is understood that $r$ is a function of the auxiliary variable $\rho$. Notice the similarity between 
this equation
and Eq.~(\ref{slTE}). In fact, the boundary condition given by Eq.~(\ref{bcgen}) assumes the same form 
here as
it does in the plane-symmetric case:
\begin{eqnarray}
\left.\left[
\overline{f^{\rm (TE)}_{\omega \ell}} \frac{d}{d\rho}{f}^{\rm (TE)}_{\omega' \ell}-
{f}^{\rm (TE)}_{\omega' \ell}\frac{d}{d\rho}\overline{f^{\rm (TE)}_{\omega \ell}}\right]\right|_{\dot{\cal I} } = 0.
\label{bcTEss}
\end{eqnarray}
This boundary condition ensures that these modes can be orthonormalized according to the sesquilinear form
given in Eq.~(\ref{sesq}), which in this spherically-symmetric scenario assumes the form
\begin{eqnarray}
\left(A, A'\right) &=& 
i \int_{\Sigma_t} d\Sigma
\left\{
\varepsilon_\parallel \bar{A}_r\partial_\tau A_r'
+\frac{\varepsilon_\bot \bar{\bf A}_\bot \cdot \partial_\tau {\bf A}'_\bot}{\gamma^2(1-n_\parallel^2 v^2)}
\right.
\nonumber \\& & 
\left. +\frac{\gamma^2(n_\parallel^2-1)v}{\mu_\bot}\left[\bar{\bf A}_\bot \cdot\partial_r{\bf A}'_\bot-(\bar{\bf A}_\bot\cdot \partial_\bot) A_r'\right]
\right\}\nonumber \\
& & 
- ({\bar{\bf A}}\leftrightarrow {\bf A}'),\;\;\;\;
\label{sesqss}
\end{eqnarray}
with $\Sigma_t$ being a spacelike surface $t = {\rm constant}$. 
After some tedious but straightforward manipulations (presented in the appendix), we obtain the 
final form of  normalized, 
positive-frequency TE modes:
\begin{eqnarray}
{\bf A}^{\rm (TE)}_{\omega \ell m} = \frac{\left(0,i m/\sin\theta,-\sin\theta \partial_\theta\right)}{\sqrt{2\omega \ell(\ell+1)}}
 e^{-i\omega \tau} Y_{\ell m}(\theta,\varphi) f_{\omega \ell }^{\rm (TE)}(r),\nonumber \\
\label{finalATEss}
\end{eqnarray}
with $f_{\omega \ell }^{\rm (TE)}$ satisfying Eqs.~(\ref{fTEss}) and (\ref{bcTEss}), and normalized according to
\begin{eqnarray}
 \int_{{\cal I}_\varrho}
d\varrho\,
\overline{f^{\rm (TE)}_{\omega \ell}} {f}^{\rm (TE)}_{\omega' \ell} = \delta_{\omega \omega'}.
\label{f2TEss}
\end{eqnarray}
Note that the integration variable is $\varrho$ [instead of $\rho$ appearing in Eq.~(\ref{fTEss})]
and ${\cal I}_\varrho$ stands for the domain of integration in this variable corresponding to ${\cal I}$
in coordinate $r$.

Now,  let us look for solutions with $A_r \nequiv 0$ --- the transverse magnetic modes, ${\bf A}^{\rm (TM)}$.
Let $\phi$ be such that $\Delta^{\!(0)}_S \phi = -r^2\varepsilon_\parallel A_r$. Thus, the gauge conditions
lead to ${\bf A}^{\rm (TM)}=(-r^{-2} \varepsilon_\parallel^{-1}\Delta^{\!(0)}_S, \partial_{\theta}\partial_\varrho,\partial_{\varphi}\partial_\varrho) \phi$. Using again stationarity and spherical symmetry, 
$\phi = e^{-i\omega \tau} Y_{\ell m}(\theta,\varphi) f_{\omega \ell }^{\rm (TM)}(r)$, we obtain that
$ f_{\omega \ell }^{\rm (TM)}(r)$ satisfies
 \begin{eqnarray}
 \left[-\frac{d^2}{d\varrho^2}+\left(\frac{\gamma^2(1-n_\parallel^2v^2)\ell(\ell+1)}
 {r^2 \varepsilon_\bot \varepsilon_\parallel}-\frac{\mu_\bot \omega^2}{\varepsilon_\bot}\right)\right] 
 f^{\rm (TM)}_{\omega \ell} = 0.\;\;\;\;
 \label{fTMss}
 \end{eqnarray}
Notice, again, the similarity between this equation and Eq.~(\ref{slTM}). And, again, the boundary condition 
imposed by Eq.~(\ref{bcgen}) to these modes take the same form as in the plane-symmetric case:
\begin{eqnarray}
\left.\left[\omega^2
\overline{f^{\rm (TM)}_{\omega \ell}} \frac{d}{d\varrho}{f}^{\rm (TM)}_{\omega' \ell}-\omega'^2
{f}^{\rm (TM)}_{\omega' \ell}\frac{d}{d\varrho}\overline{f^{\rm (TM)}_{\omega \ell}} 
\right]\right|_{\dot{\cal I}} = 0.
\label{bcTMss}
\end{eqnarray}
Properly orthonormalizing these modes using Eq.~(\ref{sesqss}) --- see appendix ---, leads to the positive-frequency TM normal modes
\begin{eqnarray}
{\bf A}^{\rm (TM)}_{\omega \ell m} &=& \frac{\left(r^{-2} \varepsilon_\parallel^{-1}\ell(\ell+1),\partial_\theta\partial_\varrho,i m\partial_\varrho\right)}{\sqrt{2\omega^3 \ell(\ell+1)}}\nonumber \\
& & \times
 e^{-i\omega \tau} Y_{\ell m}(\theta,\varphi) f_{\omega \ell }^{\rm (TM)}(r),
\label{finalATMss}
\end{eqnarray}
with $f_{\omega \ell }^{\rm (TM)}$ satisfying Eqs.~(\ref{fTMss}) and (\ref{bcTMss}), and normalized according to
\begin{eqnarray}
 \int_{{\cal I}_\rho}
d\rho\,
\overline{f^{\rm (TM)}_{\omega \ell}} {f}^{\rm (TM)}_{\omega' \ell} = \delta_{\omega \omega'}.
\label{f2TMss}
\end{eqnarray}
Similarly to the TE case, note 
that the integration variable is not the same which  appears in the differential equation,  Eq.~(\ref{fTMss}).
(${\cal I}_\rho$ stands for the domain of integration in the variable $\rho$ corresponding to ${\cal I}$
in coordinate $r$.)

The electromagnetic field operator can be represented in terms of the TE and TM modes (and their complex conjugates)
as
\begin{eqnarray}
{\hat{\bf A}} = \sum_{{\rm J}\in\{{\rm TE},{\rm TM}\}}\sum_{\ell m}
\int_{{\cal E}_{\ell  +}^{({\rm J})}} d\omega
\left[{\hat a}^{({\rm J})}_{\omega \ell m} {\bf A}^{\rm (J)}_{\omega\ell m}+ {\rm H.c.}\right],
\label{hatAss}
\end{eqnarray}
where ${\cal E}_{\ell +}^{\rm (J)} := {\cal E}_{\ell}^{\rm (J)}\cap {\mathbb R}^\ast_+$, with ${\cal E}_{\ell }^{\rm (J)} $
being the set of $\omega$ values for which Eqs.~(\ref{fTEss}) and (\ref{bcTEss}), for ${\rm J} = {\rm TE}$, and
Eqs.~(\ref{fTMss}) and (\ref{bcTMss}), for ${\rm J} = {\rm TM}$, have nontrivial solutions.
The orthonormality of TE and TM modes, 
\begin{eqnarray}
\left(A^{\rm (J)}_{\omega \ell m},A^{\rm (J')}_{\omega'\ell'm'}\right)
& =& -\left(\overline{A^{\rm (J )}_{\omega\ell m}},\overline{A^{\rm (J')}_{\omega'\ell' m'}}\right)\nonumber \\
&=&
\delta_{{\rm J}{\rm J}'}\delta_{\omega\omega'}\,\delta_{\ell \ell'} \delta_{m m'},
\label{normJJss}
\\
\left(A^{\rm (J)}_{\omega \ell m},\overline{A^{\rm (J')}_{\omega'\ell'm'}}\right) &=& 
0,
\label{normJJbarss}
\end{eqnarray}
requires that the canonical commutation relations
\begin{eqnarray}
\left[{\hat a}^{({\rm J})}_{\omega \ell m},{\hat a}^{({\rm J}')\dagger}_{\omega' \ell' m'}\right] &=& \delta^{\rm JJ'}
\delta_{\omega\omega'}\,\delta_{\ell \ell'} \delta_{m m'},
\label{CCR1ss}
\\
\left[{\hat a}^{({\rm J})}_{\omega \ell m},{\hat a}^{({\rm J}')}_{\omega' \ell' m'}\right] & =& 0
\label{CCR2ss}
\end{eqnarray}
hold.

\subsection{Instability analysis}
\label{subsec:instass}

The close similarity between Eqs.~(\ref{slTE}) and (\ref{fTEss}) and between Eqs.~(\ref{slTM})
and (\ref{fTMss}) make the instability analysis in this spherically-symmetric scenario
essentially 
identical to the one performed in the plane-symmetric case, with $\ell(\ell+1)$ playing the role
$k_\bot^2$ did  in Eq.~(\ref{Veff}). So, putting the effective potentials of Eqs.~(\ref{fTEss}) and (\ref{fTMss}),
with $\omega^2 = -\Omega^2$,
in the form
\begin{eqnarray}
V_{\text{\it eff}} = C_1 \ell(\ell+1) + C_2 \Omega^2,
\label{eq:C1C2}
\end{eqnarray}
we again have two types of instabilities: (i) large-$\ell$ instability, when $C_1 < 0$ somewhere, and (ii) minimum-thickness
instability, when $C_1 >0$ but $C_2<0$ in a sufficiently thick spherical shell --- see discussion below Eq.~(\ref{Veff}).
The only additional feature is that, by allowing the medium to flow,
type-(i)  (large-$\ell$) instability for both TE and TM modes  can arise when 
 the medium's velocity $v(r)$ exceeds the radial light velocity $n_\parallel^{-1}$.
 
 Let $g_{\Omega\ell}^{\rm (J)}$ represent the solutions of Eqs.~(\ref{fTEss}) (for ${\rm J} = {\rm TE}$) and (\ref{fTMss}) (for ${\rm J} = {\rm TM}$), subject to
 the boundary conditions given by Eqs.~(\ref{bcTEss}) and (\ref{bcTMss}), respectively, with $\omega^2 = -\Omega^2$ ($\Omega > 0$, without loss of generality).
The normalized, unstable modes are presented below --- see appendix for details.

\subsubsection{Unstable TE modes}

Unstable TE modes orthonormalized according to the analogous of Eqs.~(\ref{normJJss}) and (\ref{normJJbarss})  read (up to global phase and time translation)
\begin{eqnarray}
{\bf A}^{\rm ({\it u}TE)}_{\Omega \ell m} &=& \frac{\left( e^{\Omega \tau-i s_\varepsilon^\bot\kappa/2}+e^{-\Omega \tau+i s_\varepsilon^\bot \kappa/2}\right)}{2\sqrt{\Omega \ell(\ell+1)\sin\kappa}} g_{\Omega \ell }^{\rm (TE)}(r)
\nonumber \\
 & &\times 
\left(0,i m/\sin\theta,-\sin\theta \partial_\theta\right)
 Y_{\ell m}(\theta,\varphi),
\label{finalAuTEss}
\end{eqnarray}
with $\kappa$ being a constant ($0<\kappa<\pi$), $g_{\Omega \ell }^{\rm (TE)}$ normalized according to
\begin{eqnarray}
\left|\int_{{\cal I}} dr \frac{\varepsilon_\bot}{\gamma^2(1-n^2_\parallel v^2)}\,\overline{g^{\rm (TE)}_{\Omega \ell}}(r)\,g^{\rm (TE)}_{\Omega' \ell}(r)\right|& =&\delta_{\Omega\Omega'},
\label{guTEss}
\end{eqnarray}
and $s_\varepsilon^\bot$ being the sign of the integral above.
Calculating the electric ${\bf E}_{\Omega\ell m}^{(u{\rm TE})}$ and magnetic  ${\bf B}_{\Omega \ell m}^{(u{\rm TE})}$ 
vector fields associated to these modes in the {\it lab frame}, we have:
\begin{widetext}
\begin{eqnarray}
\E^{(u{\rm TE})}_{\Omega \ell m} &=&
 \frac{ \sqrt{\Omega}\left(-i m \,{\bf e}_\theta/\sin\theta +{\bf e}_\varphi \,\partial_\theta\right)}
 {2r\sqrt{\ell(\ell+1)\sin\kappa}}\;
g_{\Omega \ell }^{\rm (TE)}(r)Y_{\ell m}(\theta,\varphi)
 \left( e^{\Omega \tau-i s_\varepsilon^\bot\kappa/2}-e^{-\Omega \tau+i s_\varepsilon^\bot \kappa/2}\right),
\label{EuTEss}\\
\B^{(u{\rm TE})}_{\Omega \ell m}&=&\frac{\left[\ell (\ell+1)\,{\bf e}_r+\left(i m \,{\bf e}_\varphi/\sin\theta 
+{\bf e}_\theta\,
 \partial_\theta\right)r\partial_r\right]}{2r^2\sqrt{\Omega \ell(\ell+1)\sin\kappa}} 
 g_{\Omega \ell }^{\rm (TE)}(r)
 Y_{\ell m}(\theta,\varphi)
\left( e^{\Omega \tau-i s_\varepsilon^\bot\kappa/2}+e^{-\Omega \tau+i s_\varepsilon^\bot \kappa/2}\right).
\nonumber \\
\label{BuTEss}
\end{eqnarray}
\end{widetext}

\subsubsection{Unstable TM modes}

Finally, the unstable TM modes orthonormalized according to the analogous of 
Eqs.~(\ref{normJJss}) and (\ref{normJJbarss})  read (up to global phase and time translation)
\begin{eqnarray}
{\bf A}^{\rm ({\it u}TM)}_{\Omega \ell m} &=& \frac{\left(r^{-2} 
\varepsilon_\parallel^{-1}\ell(\ell+1),\partial_\theta\partial_\varrho,i m\partial_\varrho\right)}{2\sqrt{\Omega^3 
\ell(\ell+1)\sin\kappa}} g_{\Omega \ell }^{\rm (TM)}(r) \nonumber \\
& & \times Y_{\ell m}(\theta,\varphi)
\left( e^{\Omega \tau+i s_\mu^\bot\kappa/2}+e^{-\Omega \tau-i s_\mu^\bot \kappa/2}\right),\;\;\;\;\;\;
\label{finalAuTMss}
\end{eqnarray}
with, again, $\kappa$ being a constant ($0<\kappa<\pi$), $g_{\Omega \ell }^{\rm (TM)}$ normalized according to
\begin{eqnarray}
\left|\int_{{\cal I}} dr \frac{\mu_\bot}{\gamma^2(1-n^2_\parallel v^2)}\,
\overline{g^{\rm (TM)}_{\Omega \ell}}(r)\,g^{\rm (TM)}_{\Omega' \ell}(r)\right|& =&\delta_{\Omega\Omega'},\;\;\;
\label{guTMss}
\end{eqnarray}
and $s_\mu^\bot$ being the sign of the integral above.
Calculating the electric ${\bf E}_{\Omega\ell m}^{(u{\rm TM})}$ and magnetic  
${\bf B}_{\Omega \ell m}^{(u{\rm TM})}$ 
vector fields associated to these modes in the {\it lab frame}, we have:
\begin{widetext}
\begin{eqnarray}
\E^{(u{\rm TM})}_{\Omega \ell m} &=&- 
\frac{\left[\ell (\ell+1)\,{\bf e}_r/\varepsilon_\parallel+\left(i m \,{\bf e}_\varphi/\sin\theta 
+{\bf e}_\theta\,
 \partial_\theta\right)r\partial_\varrho\right]}{2r^2\sqrt{\Omega \ell(\ell+1)\sin\kappa}}  
  g_{\Omega \ell }^{\rm (TM)}(r)Y_{\ell m}(\theta,\varphi)
\left( e^{\Omega \tau+i s_\mu^\bot\kappa/2}-e^{-\Omega \tau-i s_\mu^\bot \kappa/2}\right),
\label{EuTMss}\\
\B^{(u{\rm TM})}_{\Omega \ell m}&=&\frac{\mu_\bot\sqrt{\Omega}\left(-im\,{\bf e}_\theta/\sin\theta+
{\bf e}_\varphi\,
\partial_\theta
\right)}{2r
\gamma^2(1-n^2_\parallel v^2)\sqrt{
\ell(\ell+1)\sin\kappa}} g_{\Omega \ell }^{\rm (TM)}(r) 
Y_{\ell m}(\theta,\varphi)
\left( e^{\Omega \tau+i s_\mu^\bot\kappa/2}+e^{-\Omega \tau-i s_\mu^\bot \kappa/2}\right).
\label{BuTMss}
\end{eqnarray}
\end{widetext}

\vskip 0.5cm

As argued in the previous case, when instability is triggered and modes ${\bf A}^{\rm ({\it u}
J)}_{\Omega \ell m}$
appear, they must be included in the field expansion given by Eq.~(\ref{hatAss}), along with their 
complex conjugates. Eventually ($t\gg \Omega^{-1}$), these  modes  dominate the field fluctuations.

\subsection{Example}
\label{subsec:exss}

\begin{figure}[t]
\includegraphics[scale=0.27]{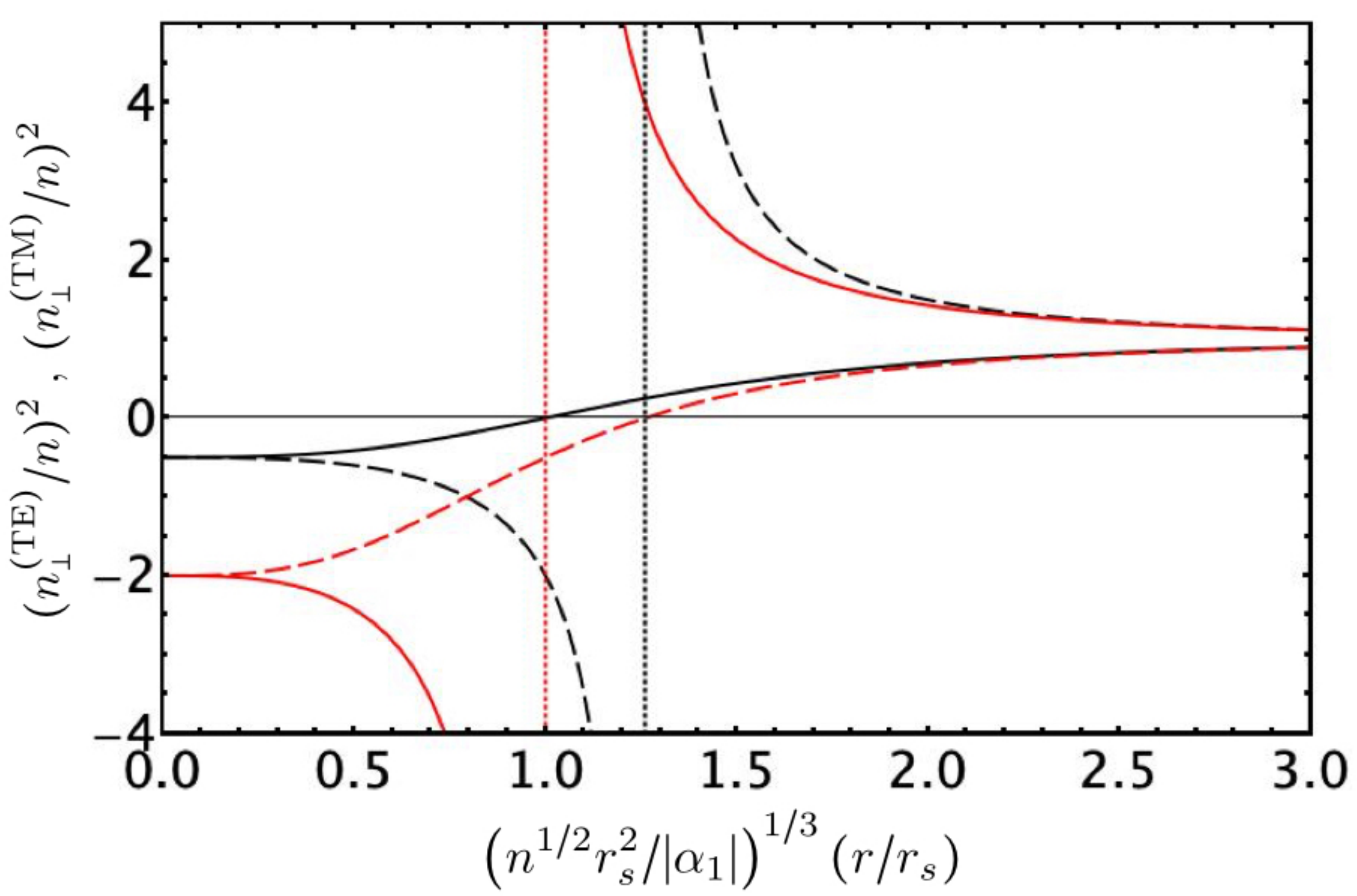}
\caption{Squared values of the refractive indices $n_\bot^{\rm (TE)}$ (in black) and  $n_\bot^{\rm (TM)}$ (in red) for positive (solid lines) and negative (dashed lines)
values of $\alpha_1$. The black and red dotted lines mark where $n_\bot^{\rm (TE)}$ (for negative $\alpha_1$) and $n_\bot^{\rm (TM)}$ (for positive $\alpha_1$) 
are singular, respectively.}
\label{fig:n}
\end{figure}

Now, let us consider a concrete scenario where electromagnetism in a gravitationally interesting system,
nonminimally coupled to the background geometry 
via $\chi^{abcd}$ given by Eq.~(\ref{QED}) (but with arbitrary 
$\alpha_1,\alpha_2,\alpha_3$), can be
 mimicked by an anisotropic, stationary moving medium. 
 We have already seen that
 setting $n = {\rm constant}$ and $v^2 = [1+(n^2-1)r/r_s]^{-1}$, leads to an effective line element which describes
 the
 vacuum Schwarzschild spacetime. In this case, Eq.~(\ref{constr}) is trivially satisfied 
 and Eqs.~(\ref{musol}--\ref{chimusol}) give
 \begin{eqnarray}
 \mu &=& n,
 \label{ssmusol}
 \\
 \Delta^{(\varepsilon)} & = &3\alpha_1 n^{1/2} \frac{r_s}{r^3} ,
 \label{ssDeltaepssol}
 \\
 \Delta^{(\mu)} & = &\frac{3\alpha_1}{n^{3/2}} \frac{r_s}{r^3} ,
 \label{ssDeltamusol}
 \end{eqnarray}
which lead to the material properties
 \begin{eqnarray}
 \varepsilon_\bot &=& n\left(1-\frac{\alpha_1 r_s}{n^{1/2}r^3}\right),
 \label{ebotS}
 \\
 \varepsilon_\parallel & = &n\left(1+\frac{2\alpha_1 r_s}{n^{1/2}r^3}\right) ,
 \label{eparallelS}
 \\
\mu_\bot & = &n\left(1-\frac{\alpha_1 r_s}{n^{1/2}r^3}\right)^{-1},
 \label{mubotS}
 \\
\mu_\parallel & = &n\left(1+\frac{2\alpha_1 r_s}{n^{1/2}r^3}\right)^{-1} .
 \label{muparallelS}
 \end{eqnarray}
 We promptly see that $n_\parallel := \sqrt{\mu_\bot\varepsilon_\bot} = n$, which shows that the analogue horizon for these nonminimally-coupled modes,  located 
 where $v^2 = n_\parallel^{-2}$,
 coincides with the analogue Schwarzschild radius $r_s$. [Note, however, that this system is analogous to a physical black hole with Schwarzschild radius
 $R_s = n^{1/4} r_s$, due to absorption of $\sqrt{n}$ in Eq.~(\ref{sec4eq6}).]
 As for the other refractive indices, $n_\bot^{\rm(TE)}:=\sqrt{\mu_\parallel\varepsilon_\bot}$ and $n_\bot^{\rm(TM)}:=\sqrt{\mu_\bot\varepsilon_\parallel} $ ($=n^2/n^{\rm (TE)}_\bot$),
 Fig.~\ref{fig:n} shows their squared values  (in black and red, respectively) for positive (solid lines) and 
 negative (dashed lines) values of $\alpha_1$.
 Note that, depending on the values of $\alpha_1/(n^{1/2}r_s^2)$, some kind of metamaterial (possibly with some negative squared refractive indices) 
 may be needed in order to mimic this nonminimal
coupling of the electromagnetic field with the Riemann curvature tensor in the exterior region of a Schwarzschild black hole. Conversely,
regardless how difficult it may be to set up such an experimental configuration in the lab, 
it is interesting in its own  that 
QED-inspired nonminimally-coupled electromagnetism in the background of a black hole behaves as  in
such an exotic metamaterial in flat spacetime.
\begin{figure*}
 \includegraphics[width=\textwidth,height=13cm]{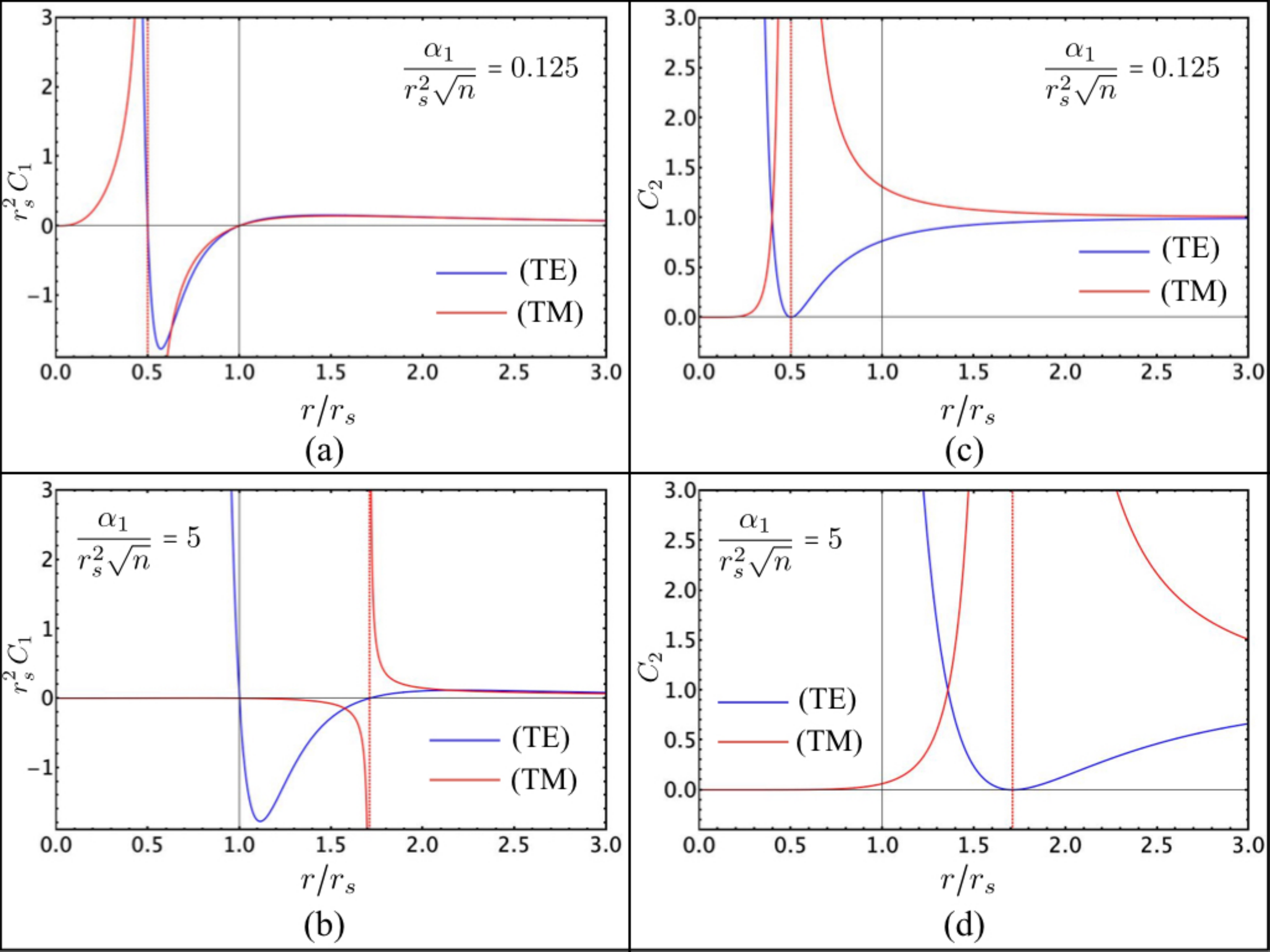}
\caption{Plot of the coefficients $C_1$ --- (a) and (b) --- and  $C_2$ --- (c) and (d) ---
appearing in Eq.~(\ref{eq:C1C2}) for electromagnetic modes
TE (blue curves) and TM (red curves),
nonminimally coupled to the background geometry of a Schwarzschild black hole via
 Eq.~(\ref{QED}). Figs.~(a) and (c) illustrate the general behavior of $C_1$ and $C_2$
 for $-r_s^2\sqrt{n}/2<\alpha_1 <r_s^2\sqrt{n}$, while
 (b) and (d) are representative of the 
 behavior of $C_1$ and $C_2$ for $\alpha_1<-r_s^2\sqrt{n}/2$ or $\alpha_1 > r_s^2\sqrt{n}$. According to the
 instability discussion, only large-$\ell$ instability can appear in this case, since $C_2\geq 0$ everywhere. Moreover, 
 for $\alpha_1< - r_s^2\sqrt{n}/2$ or $\alpha_1 > r_s^2\sqrt{n}$, the unstable modes can be mostly supported outside
 the analogous event horizon, $r >r_s$.}
\label{fig:Vs}
\end{figure*}

Turning to the question of possible instabilities, in Fig.~\ref{fig:Vs} we show the behavior of the terms $C_1$ and $C_2$ appearing in Eq.~(\ref{eq:C1C2})
for the TE (in blue) and TM (in red) modes --- extracted, respectively, from Eqs.~(\ref{fTEss}) and (\ref{fTMss}):
\begin{eqnarray}
C_1 &=& 
\left\{
\begin{array}{l}
n^{-2}r^{-9}\left(r-r_s\right)\left(r^3-\frac{\alpha_1r_s}{\sqrt{n}}\right)\left(r^3+\frac{2\alpha_1r_s}{\sqrt{n}}\right)\\
\frac{n^{-2}\left(r-r_s\right)r^3}{\left(r^3-\alpha_1r_s/\sqrt{n}\right)\left(r^3+2\alpha_1r_s/\sqrt{n}\right)}
\end{array}
\right.,\;\;\;\;\;
\label{C1ss}
\\
C_2 &=& 
\left\{
\begin{array}{l}
\left(1-\frac{\alpha_1r_s}{r^3\sqrt{n}}\right)^2\\
\left(1-\frac{\alpha_1r_s}{r^3\sqrt{n}}\right)^{-2}
\end{array}
\right.,\;\;\;
\label{C2ss}
\end{eqnarray}
where the first and second lines in the expressions above refer to the TE and TM modes,
respectively. The  Fig.~\ref{fig:Vs}(a) 
is representative of the behavior of $C_1$ for $-r_s^2\sqrt{n}/2<\alpha_1<r_s^2\sqrt{n}$, 
while Fig.~\ref{fig:Vs}(b) gives the correct qualitative
behavior of $C_1$ for $\alpha_1<-r_s^2\sqrt{n}/2$ or $\alpha_1>r_s^2\sqrt{n}$. Figs.~\ref{fig:Vs}(c) and
\ref{fig:Vs}(d) show the behavior of $C_2$
for the same values of $\alpha_1$ used in Figs.~\ref{fig:Vs}(a) 
and
\ref{fig:Vs}(b).

It is clear, from the expressions above, that $C_2$ is everywhere non-negative, while $C_1$ assumes negative 
values in the region with radial coordinate $r$
between $(\alpha_1 r_s/\sqrt{n})^{1/3}$ and $r_s$ (if $\alpha_1>0$) or between
$[|\alpha_1| r_s/(2\sqrt{n})]^{1/3}$ and $r_s$ (if $\alpha_1<0$). Therefore, according to the discussion
of Subsec.~\ref{subsec:instass}, this nonminimally-coupled electromagnetic theory in Schwarzschild
spacetime exhibits
large-$\ell$ instability. In particular, if $\alpha_1>r_s^2\sqrt{n}$ or $\alpha_1<-2 r_s^2\sqrt{n}$, then the unstable
modes influence the exterior region of the back hole.


\section{Stabilization: spontaneous vectorization, photo production, and long-range induced correlations}
\label{sec:stabilization}

We now turn our attention to discussing what can possibly happen to the analogous system when the  
vacuum instability is triggered. In the gravitational scenario, it has been shown that in some cases
(for instance, depending on the field-background coupling),
stabilization occurs due to the appearance of a nonzero value for the field (spontaneous 
scalarization/vectorization)~\cite{PCBRS,DE1,DE2,B,Cardoso2019},
seeded by decoherence of the growing initial-vacuum fluctuations~\cite{LLMV}.
In this process, field particles/waves are produced~\cite{PCBRS,LLMV2} 
and carry away the energy excess of the initial
vacuum state in comparison to the stabilized configuration.

If we  transpose these conclusions, {\it mutatis mutandis}, to our analogous systems,
then an electromagnetic field should spontaneously appear in the material, bringing
the whole system to a new equilibrium configuration --- through
nonlinear effects brought in by field-dependent constitutive tensors
 $\varepsilon^{ab}$ and $\mu_{ab}$ [see Eqs.~(\ref{DE},\ref{HB})] ---, with photons being emitted, carrying away the energy excess. 
Although the detailed dynamics of the stabilization processes in the gravitational and  in the analogous systems
are quite different  --- ruled by Einstein equations in the gravitational case and by the macroscopic
Maxwell's equations with field-dependent
 $\varepsilon^{ab}$ and $\mu_{ab}$ 
in the analogous systems ---, the qualitative features of the whole process, described above,
seem quite reasonable to occur in generic field stabilization processes.

It is important to mention that the time scale set by the instability, $\Omega^{-1}$, is typically of the order of the
time light takes to travel the typical size of the system, $L$. Therefore, in the analogous lab 
scenarios, the stabilization process
would occur
almost instantaneously ($\sim L/(1~{\rm cm}) \times 10^{-10}~{\rm s}$) once the instability conditions are met
--- which, for a given system, may depend on external parameters such as temperature, external fields, etc.,
through their influence on the constitutive functions $\varepsilon_\bot$, $\varepsilon_\parallel$,
$\mu_\bot$, $\mu_\parallel$. The whole process would most likely be interpreted as a kind of phase transition, 
where the ``long-range'' emergent correlations in the material would come from interaction of its constituents with 
a common (initially-unstable vacuum) fluctuating mode and/or the stabilized field configuration.

For concreteness sake, let us consider the explict form of the unstable modes found in the example of Sec.~\ref{sec:ani},
where instability occurs due to a negative value of $\mu_\bot$ --- for some (isolated) $\omega^2 = -\Omega^2 <0$ ---
in a homogeneous slab of width $L$.
Although this system is not analogous to vacuum nonminimally-coupled electromagnetism in any realistic
spacetime, it serves to illustrate  general features  of the mechanism itself, in addition to being much simpler
to setup in the lab. This is no different than
looking for fingerprints of analogue Hawking radiation in systems whose only similarity with realistic black holes
is the presence of an effective event horizon --- which is the common approach in condensed-matter and optical 
experimental analogues.

\begin{figure*}
 \includegraphics[width=\textwidth,height=13cm]{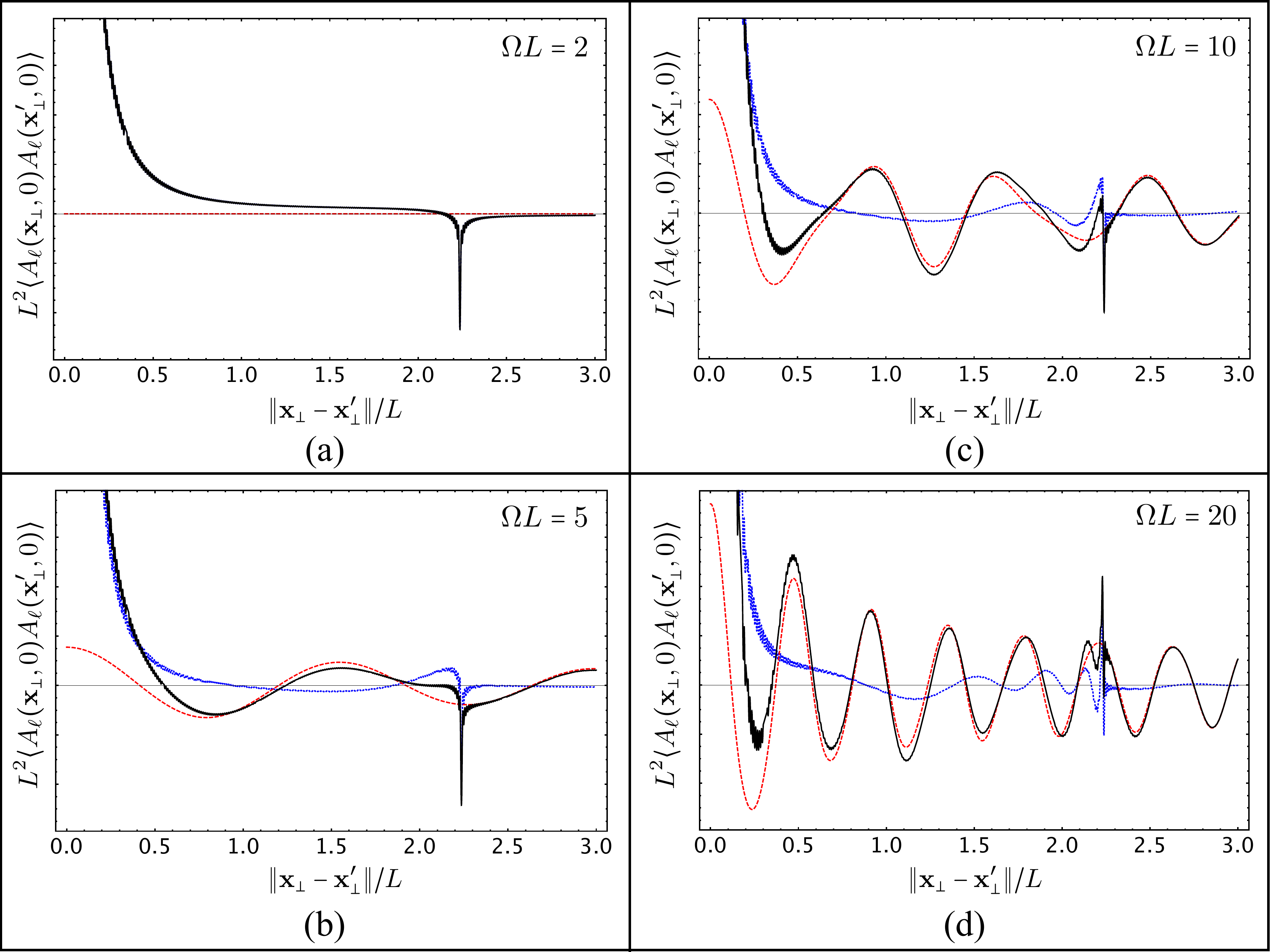}
\caption{Equal-time ($t=t'\gg \Omega^{-1}$), two-point correlation function $\langle { A}_\ell({\bf x}) { A}_\ell({\bf x}')\rangle$ of the component of the quantum field $\hat{{\bf A}}$ along the vector-separation
${\bf x}_\bot-{\bf x}_\bot'$, for points in the $z=0$ plane, for  different values of $\Omega L$ --- with same values of constitutive functions given in Fig.~\ref{fig:numsol} . The dotted (blue) lines represent
the contribution coming from the TE modes, while the dashed (red) lines depict the 
contribution coming from the TM modes. 
The solid (black) lines give the sum of both contributions. Notice that
long-range ($\|{\bf x}_\bot-{\bf x}_\bot'\| \gtrsim L$) correlations are mainly due to the TM modes, which undergo
minimum-width instability.}
\label{fig:CorrCenter}
\end{figure*}

As argued before,
once instability sets in, the unstable modes must be added to the expansion of the field operator $\hat{\bf A}$, along with their complex 
conjugates, 
with corresponding
annihilation $\hat{a}^{(u{\rm J})}_{\Omega \k_\bot}$ and creation $\hat{a}^{(u{\rm J})\dagger}_{\Omega \k_\bot}$
operators. It is easy to see that the field's vacuum fluctuations and correlations are eventually ($t,t'\gg \Omega^{-1}$) dominated
by  these
unstable modes --- at least as long as decoherence does not come into play. The dominant contribution to the vacuum 
correlations in the example of
Subsec.~\ref{subsec:explane} reads
(the reader should refer  
to Subsec.~\ref{subsec:explane} for the definition of all quantities appearing in these expressions):
\begin{widetext}
\begin{eqnarray}
\langle { A}_j({x}) { A}_l({x}')\rangle \!\!&\sim&\!\!\!\! \frac{2\pi}{L_\bot}\int_0^{2\pi}\!\! \!\!\!\!\!\!\!d\varphi\left\{\!\!\!\!\!\!\!\!\sum_{m=0}^{\;\;\;\;\;\;m^{\text{\tiny{(TM)}}}}
\!\!\! k_\bot^{(m)}
\left[{\rm \A}^{\rm (uTM)}_{\Omega \k^{(m)}_\bot}({x})\right]_j \left[\overline{{\rm \A}^{\rm (uTM)}_{\Omega \k_\bot^{(m)}}}({x}')\right]_l+
\!\!\!\!\!\!\!\!\!\sum_{\;\;\;\;\;\;m=m^{\text{\tiny{(TE)}}}}^\infty
\!\! k_\bot^{(m)}
\left[{\rm \A}^{\rm (uTE)}_{\Omega \k_\bot^{(m)}}({x})\right]_j \left[\overline{{\rm \A}^{\rm (uTE)}_{\Omega \k_\bot^{(m)}}}({ x}')\right]_l\right\}
\nonumber \\
& &\!\!\!\!\!\!\!\!\!\!\!\!\!\!\!\!\!\!\! \!\!\!\!\!\sim \frac{e^{\Omega(t+t')}}{4 L_\bot\Omega}\left\{\!\!\!\!\!\!\!\!\sum_{m=0}^{\;\;\;\;\;\;m^{\text{\tiny{(TM)}}}}
\!\!\! \frac{k_\bot^{(m)}}{\sin\kappa_m}
{\cal D}^{\rm (TM)}_{jl(m)}(d_\bot)g^{\rm (TM)}_{\Omega k^{(m)}_\bot}(z){\overline{g^{\rm (TM)}_{\Omega k^{(m)}_\bot}}}(z')+
\!\!\!\!\!\!\!\!\!\sum_{\;\;\;\;\;\;m=m^{\text{\tiny{(TE)}}}}^\infty
\!\! \frac{k_\bot^{(m)}}{\sin\kappa_m}
{\cal D}^{\rm (TE)}_{jl(m)}(d_\bot)g^{\rm (TE)}_{\Omega k^{(m)}_\bot}(z){\overline{g^{\rm (TE)}_{\Omega k^{(m)}_\bot}}}(z')\right\},\;\;\;\;\;\;\;\;\;\;
\label{AAuTE}
\end{eqnarray}
where $\varphi$ is the angle between ${\bf k}_\bot$ and $({\bf x}_\bot-{\bf x}'_\bot)$,
 $d_\bot := \|{\bf x}_\bot - {\bf x}_\bot'\|$,
and the operators ${\cal D}^{\rm (J)}_{jl(m)}(d_\bot)$ acting on $g^{\rm (J)}_{\Omega k^{(m)}_\bot}(z){\overline{g^{\rm (J)}_{\Omega k^{(m)}_\bot}}}(z')$ are defined by
\begin{eqnarray}	
{\cal D}^{\rm (TM)}_{jl(m)}(d_\bot)&:=&\frac{1}{\Omega^2\varepsilon_\bot^2}\left[J'_1\left(k_\bot^{(m)}d_\bot\right)\delta^\ell_j \delta^\ell_l+
\frac{J_1\left(k_\bot^{(m)}d_\bot\right)}{k_\bot^{(m)}d_\bot}\delta^\varphi_j \delta^\varphi_l\right]\frac{d^2}{dzdz'}+
\frac{\left(k^{(m)}_\bot\right)^2J_0\left(k_\bot^{(m)}d_\bot\right)}{\Omega^2\varepsilon_\parallel^2}\delta^z_j \delta^z_l
\nonumber \\
& &
-\frac{k^{(m)}_\bot J_1\left(k_\bot^{(m)}d_\bot\right)}{\Omega^2\varepsilon_\parallel \varepsilon_\bot}
\left(\delta^\ell_j \delta^z_l\frac{d}{dz}-\delta^z_j \delta^\ell_l\frac{d}{dz'}\right),
\label{calDTM}
\\
{\cal D}^{\rm (TE)}_{jl(m)}(d_\bot)&:=&\frac{J_1\left(k_\bot^{(m)}d_\bot\right)}{k_\bot^{(m)}d_\bot}\delta^\ell_j \delta^\ell_l+J'_1\left(k_\bot^{(m)}d_\bot\right)\delta^\varphi_j \delta^\varphi_l,
\label{calDTE}
\end{eqnarray}
\end{widetext}
with indices $\ell$ and $\varphi$ standing for vector components along $({\bf x}_\bot-{\bf x}'_\bot)$
and ${\bf n}_z\times({\bf x}_\bot-{\bf x}'_\bot)$, respectively; $J_n$ and $J_n'$ 
stand for the Bessel functions of first kind
and their first derivatives, respectively.
Field correlations $\langle { E}_j({x}) {E}_l({x}')\rangle $ and $\langle { B}_j({x}) { B}_l({x}')\rangle $
can be similarly obtained --- in particular,
$\langle { E}_j({x}) {E}_l({x}')\rangle \sim \Omega^2\langle { A}_j({x}) { A}_l({x}')\rangle$. 
As an illustration, 
in Fig.~\ref{fig:CorrCenter} we plot the {\it equal-time} ($t=t'\gg \Omega^{-1}$), 
{\it longitudinal} correlation function
$\langle { A}_\ell({\bf x}) { A}_\ell({\bf x}')\rangle $ for points ${\bf x}$, ${\bf x}'$ in the plane $z = 0$, for the same values
of constitutive functions used in Fig.~\ref{fig:numsol} and four different values of $\Omega L$. The vertical-axis 
 scale
is arbitrary --- but the same in all plots ---, since the correlations 
grow exponentially in time, from their typical (stable-vacuum) values of order
 $\hbar/(cL d_\bot)\sim [1~{\rm cm}^2/(L d_\bot)]\times10^{-8}~{\rm eV}/({\rm cm}^3\, {\rm GHz}^2)$, until
 decoherence and vectorization take over.
Notice that once minimum-width (TM) instability sets in, macroscopic ($\sim L$) field correlations
are enhanced. It is an interesting question whether any such ``long-range'' correlation would survive or leave
an inprint in the final stable configuration. Although not directly relevant for the analogy with gravity-induced instability
 itself, such correlations might lead to interesting material behavior.



\section{Final Remarks}
\label{sec:final}

We have shown that gravity-induced instabilities, related to the vacuum-awakening effect in the quantum 
context~\cite{LV,LMV,LMMV,LLMV2}
 and
spontaneous scalarization/vectorization in the classical one~\cite{PCBRS,DE1,DE2,B,Cardoso2019}, 
can be mimicked by electromagnetism in anisotropic 
metamaterials with appropriate constitutive functions. This follows from the formal analogy between electromagnetism
in anisotropic media and nonminimally-coupled electromagnetism 
in curved spacetimes, presented in Sec.~\ref{sec:cov}. 
We explored two concrete scenarios: (i) a plane-symmetric, static slab ---
whose main interest is its simplicity regarding experimental setup (see Sec.~\ref{sec:ani}) --- and (ii)
a spherically-symmetric, moving media --- whose main feature is its analogy with QED-inspired
nonminimally-coupled electromagnetism in Schwarzschild spacetime~\cite{DH,Lemos} for given velocity
and constitutive-functions profiles (see Subsec.~\ref{subsec:QED} and Sec.~\ref{sec:spherical}).

Once instability is triggered in the analogous systems, some stabilization process must take place, leading
the system to a new stable configuration. The details of this stabilization process and of the final configuration
will most likely  depend on specific nonlinear properties of the metamaterial, but it seems reasonable that they
might involve the appearance of nonzero electromagnetic fields in the material (analogous to
spontaneous vectorization in curved spacetimes) 
and  photo production  which carries away the energy excess with respect to the
stable configuration. As discussed earlier, the time scale involved in the stabilization process can be very short
($\sim 10^{-10}$~s), which would make it very difficult to even identify the unstable phase.
This is similar to what might occur with negative conductivity, which has never been
directly measured but which is 
predicted to lead to 
zero-dc-resistance states~\cite{AAM2003} which were observed in laboratory~\cite{MSKNJU,ZDPW}
--- although an alternative explanation has been proposed~\cite{IP}.

Clearly, the feasibility of such analogues is bound to 
the existence of material configurations with the required constitutive functions. As briefly pointed out in the introduction, this can be achieved at least for anisotropic neutral plasmas, and the recent advances in metamaterial science offer a plethora of possible candidates, specially the hyperbolic metamaterials~\cite{Poddubny2013,Caligiuri2016}, that possess precisely the form given in
Eqs.~(\ref{epssplit},\ref{musplit}) with the required ``negativeness.'' In particular, we call 
attention to the increase in the spontaneous light emission in such configurations, which 
may be related to the process of stabilization in active scenarios.

It is also important to mention that the QED-inspired analogues (Subsec.~\ref{subsec:QED}) 
are not restricted to the study of vacuum instability. For instance, they can be used to study light ray propagation in the corresponding spacetimes and one possible application is the QED-induced birefringence in the Schwarzschild spacetime~\cite{DH}. For this particular experiment, one can work far from 
the effective horizon, where the constitutive coefficients \eqref{ebotS}--\eqref{muparallelS} are positive.

Our main purpose here  
was to lay down a novel class of analogue models of  curved-spacetime phenomena, with main interest on 
the gravitational side of the analogy. Notwithstanding, the consequences of the analogue 
gravity-induced instability
to the metamaterial side may be interesting on its own. The electromagnetic field instability may mark, lead or
mediate some kind of phase transition in the
metamaterial, where the spontaneously created field and/or its 
amplified ``long-range'' correlations may play some important
role (see discussion in Sec.~\ref{sec:stabilization}).
Investigation in these lines are currently in course and will be presented elsewhere.

\acknowledgments
C.\ R.\ 
was mainly supported
by S\~ao Paulo  Research Foundation (FAPESP), through Grant No.\ 2015/26438-8, and in part by the 
Coordena\c c\~ao de Aperfei\c coamento de Pessoal de N\'ivel Superior - Brasil (CAPES) - Finance Code 001.

\appendix
\section{Normalization of stable and unstable modes in the spherically symmetric case}
\label{ap:ss}

Here, we present in detail the calculations involved in normalizing the electromagnetic modes in the
spherically symmetric case. Since we are dealing with analogues to which there is a natural physical notion of time --- the lab-frame time $t$ ---, it is convenient  to use
$t = {\rm constant}$ surfaces ($\Sigma_t$)
to normalize the modes. Obviously, this choice bears no physical consequence on our results.
 
 The sesquilinear form given in Eq.~(\ref{sesq}), applied to
the scenario described in Sec.~\ref{sec:spherical}, takes the form --- notice that the integrand is a scalar and, as such, can be evaluated in any coordinate system:
\begin{eqnarray}
\left(A, A'\right) &=& 
i \int_{\Sigma_t} d\Sigma
\left\{
\varepsilon_\parallel \bar{A}_r\partial_\tau A_r'
+\frac{\varepsilon_\bot \bar{\bf A}_\bot \cdot \partial_\tau {\bf A}'_\bot}{\gamma^2(1-n_\parallel^2 v^2)}
\right.
\nonumber \\& & 
\left. +\frac{\gamma^2(n_\parallel^2-1)v}{\mu_\bot}\left[\bar{\bf A}_\bot \cdot\partial_r{\bf A}'_\bot-(\bar{\bf A}_\bot\cdot \partial_\bot) A_r'\right]
\right\}\nonumber \\
& & 
- ({\bar{\bf A}}\leftrightarrow {\bf A}').\;\;\;\;
\label{sesqss}
\end{eqnarray}
Below, we evaluate this expression for each type of mode.

\subsection{TE modes}
\label{subsec:TE}

\subsubsection{Stable}
\label{subsubsubsec:sTE}

Substituting ${\bf A}^{\rm (TE)} = (0,\partial_\varphi\psi/\sin\theta,-\sin\theta \partial_\theta\psi)$ into
 Eq.~(\ref{sesqss}), with $\psi = e^{-i\omega\tau}Y_{\ell m}(\theta,\varphi)f_{\omega \ell}^{\rm (TE)}(r)$, one gets:
\begin{eqnarray}
\left(A^{\rm (TE)}, A'^{\rm (TE)}\right) =
 \int_{S^2} dS\, \left[
(\partial_\theta \overline{Y_{\ell m}})( \partial_\theta{Y}_{\ell' m'})
\frac{}{}
\right.\nonumber \\
\left.
\frac{}{}
+\frac{m m'}{\sin^2\theta}\overline{Y_{\ell m}} {Y}_{\ell' m'}\right]\nonumber \\
\times
\int_{\cal I}d\rho\,e^{i(\omega-\omega')\tau}
\left[
(\omega+\omega')\frac{\varepsilon_\bot}{\mu_\bot} \overline{f^{\rm (TE)}_{\omega \ell}} {f}^{\rm (TE)}_{\omega' \ell'}\right.
\nonumber \\
\left.
+\frac{i \gamma^2(n_\parallel^2-1) v}{\mu_\bot}\left(
\overline{f^{\rm (TE)}_{\omega \ell}} \frac{d}{d\rho}{f}^{\rm (TE)}_{\omega' \ell'}-
{f}^{\rm (TE)}_{\omega' \ell'}\frac{d}{d\rho}\overline{f^{\rm (TE)}_{\omega \ell}} 
\right)
\right],\;\;
\label{}
\end{eqnarray}
where $S^2$ is the unit sphere, recall that $dr/d\rho = \gamma^2(1-n_\parallel^2v^2)/\mu_\bot$, and
it is understood that this last integral must be evaluated at
$ \tau +p(r) =t =  {\rm constant}$ [recall definition of $\tau$ right above Eq.~(\ref{dpdr})].
It is straightforward to show that the first integral evaluates to $\ell (\ell+1)
\delta_{\ell \ell'}\delta_{m m'}$
provided we normalize $Y_{\ell m}$ according to $\int_{S^2}dS \,\overline{Y_{\ell m}}Y_{\ell' m'}
=\delta_{\ell \ell'}\delta_{m m'}$. As for the second integral, let us first consider the quantity 
\begin{eqnarray}
W^{(\ell)}_{\omega \omega'} :=\frac{1}{(\omega-\omega')}\left(
\overline{f^{\rm (TE)}_{\omega \ell}} \frac{d}{d\rho}{f}^{\rm (TE)}_{\omega' \ell}-
{f}^{\rm (TE)}_{\omega' \ell}\frac{d}{d\rho}\overline{f^{\rm (TE)}_{\omega \ell}}\right).\;
\label{Wdf}
\end{eqnarray}
Making use of Eq.~(\ref{fTEss}), $W^{(\ell)}_{\omega \omega'}$ clearly satisfies
\begin{eqnarray}
\frac{d}{d\rho}W^{(\ell)}_{\omega \omega'} = \frac{\varepsilon_\bot}{\mu_\bot}(\omega+\omega')
 \overline{f^{\rm (TE)}_{\omega \ell}} {f}^{\rm (TE)}_{\omega' \ell}.
 \label{dWdrho}
\end{eqnarray}
Therefore, 
\begin{eqnarray}
& & \left(A^{\rm (TE)}, A'^{\rm (TE)}\right) =
\ell (\ell+1)\delta_{\ell \ell'} \delta_{m m'}\nonumber \\
& & \times
\int_{\cal I}d\rho\,e^{i(\omega-\omega')\tau}
\left[
\frac{d}{d\rho}W^{(\ell)}_{\omega \omega'}-i(\omega-\omega')\frac{dp}{d\rho}
W^{(\ell)}_{\omega \omega'}
\right]\nonumber \\
& & = \ell (\ell+1)\delta_{\ell \ell'} \delta_{m m'}\nonumber \\
& & \times
\int_{\cal I}d\rho\,e^{i(\omega-\omega')[\tau+p(r)]}
\frac{d}{d\rho}\left(e^{-i(\omega -\omega')p(r)}W^{(\ell)}_{\omega \omega'}\right)
\nonumber \\
& & = \ell (\ell+1)\delta_{\ell \ell'} \delta_{m m'} e^{i(\omega-\omega')t}
\left.\left[e^{-i(\omega -\omega')p(r)}W^{(\ell)}_{\omega \omega'}\right]\right|_{\dot{\cal I}},\;\;\;\;\;\;
\label{AAWTE}
\end{eqnarray}
where we made use that $t = \tau+p(r)$ is kept constant 
along integration in $r$ (or $\rho$) and $[ ]|_{\dot{\cal I}}$ indicates that we must calculate the flux of the quantity in
 square brackets at the
boundaries of ${\cal I}$. 
We see that in order to guarantee orthogonality between modes with different $\omega$, without 
worring about the specific form of $p(r)$, we must impose boundary conditions at $\dot{\cal I}$ such
that,
in Eq.~(\ref{Wdf}), $\left.W^{(\ell)}_{\omega \omega'}\right|_{\dot{\cal I}}=0$ for $\omega \neq \omega'$.
Then, referring back to Eq.~(\ref{dWdrho}) and writing
\begin{eqnarray}
W^{(\ell)}_{\omega \omega'} (\rho)=(\omega+\omega') \int_{\rho_-}^{\rho}d\rho' \frac{\varepsilon_\bot}{\mu_\bot}
\overline{f^{\rm (TE)}_{\omega \ell}} {f}^{\rm (TE)}_{\omega' \ell},
\label{Wint}
\end{eqnarray}
we finally obtain
\begin{eqnarray}
 \left(A^{\rm (TE)}, A'^{\rm (TE)}\right)  =  2\omega
\ell (\ell+1) \delta_{\ell \ell'} \delta_{m m'}
 \int_{\cal I}
d\varrho\,
\overline{f^{\rm (TE)}_{\omega \ell}} {f}^{\rm (TE)}_{\omega' \ell},\nonumber \\
\label{normalizacaossTE}
\end{eqnarray}
which justifies the normalization of the TE modes in Sec.~\ref{sec:spherical}.
(Notice that the integration variable is $\varrho$, defined through $dr/d\varrho
=\gamma^2(1-n_\parallel^2v^2)/\varepsilon_\bot$.)

\subsubsection{Unstable TE modes}
\label{subsubsec:uTE}

Generic unstable TE modes are given by 
${\bf A}^{\rm ({\it u}TE)} = (0,\partial_\varphi\psi/\sin\theta,-\sin\theta \partial_\theta\psi)$  with
\begin{eqnarray}
\psi =(\alpha_{\Omega \ell} e^{\Omega\tau} + \beta_{\Omega \ell} e^{-\Omega\tau} )Y_{\ell m}(\theta,\varphi)g_{\Omega \ell}^{\rm (TE)}(r),
\label{generalUTE}
\end{eqnarray}
where $\alpha_{\Omega \ell}$ and $\beta_{\Omega \ell}$ are complex constants and $g_{\Omega \ell}^{\rm (TE)}(r)$ is a solution of Eq.~(\ref{fTEss}) with $\omega^2 = -\Omega^2$
($\Omega >0$, without loss of generality) and proper boundary conditions
(see below). Sesquilinearity of Eq.~(\ref{sesqss}) makes it easy to calculate $ \left(A^{\rm ({\it u}TE)}, A'^{\rm ({\it u}TE)}\right) $ from Eq.~(\ref{AAWTE}) with the appropriate substitution
$\omega \mapsto \mp i\Omega$ and $\omega'\mapsto \pm i\Omega'$:
\begin{widetext}
\begin{eqnarray}
& & \left(A^{\rm ({\it u}TE)}, A'^{\rm ({\it u}TE)}\right) 
= \ell (\ell+1)\delta_{\ell \ell'} \delta_{m m'} 
\nonumber \\
& &\hskip 0.5cm \times\left. \left[\overline{\alpha_{\Omega \ell}} \alpha_{\Omega'\ell}e^{(\Omega+\Omega')\tau}
W^{(u \ell)}_{\Omega \Omega'}+\overline{\beta_{\Omega \ell}} \beta_{\Omega'\ell}e^{-(\Omega+\Omega')\tau}
W^{(u \ell)}_{-\Omega\; - \Omega'}+\overline{\alpha_{\Omega \ell}} \beta_{\Omega'\ell}e^{(\Omega-\Omega')\tau}
W^{(u \ell)}_{\Omega\;- \Omega'}+\overline{\beta_{\Omega \ell}} \alpha_{\Omega'\ell}e^{-(\Omega-\Omega')\tau}
W^{(u \ell)}_{-\Omega \Omega'}\right]\right|_{\dot{\cal I}},\;\;\;\;\;\;\;\;\;
\label{AAWuTE}
\end{eqnarray}
\end{widetext}
where
\begin{eqnarray}
W^{(u\ell)}_{\pm\Omega \,\pm\Omega'} &:=&\frac{i}{(\pm\Omega\pm\Omega')}\left(
\overline{g^{\rm (TE)}_{\Omega \ell}} \frac{d}{d\rho}{g}^{\rm (TE)}_{\Omega' \ell}-
{g}^{\rm (TE)}_{\Omega' \ell}\frac{d}{d\rho}\overline{g^{\rm (TE)}_{\Omega \ell}}\right).
\nonumber \\
\label{Wudf}
\end{eqnarray}
As with the stable case, we must impose boundary conditions on $g_{\Omega \ell}^{\rm (TE)}(r)$ such that 
$$
\left.\left(
\overline{g^{\rm (TE)}_{\Omega \ell}} \frac{d}{d\rho}{g}^{\rm (TE)}_{\Omega' \ell}-
{g}^{\rm (TE)}_{\Omega' \ell}\frac{d}{d\rho}\overline{g^{\rm (TE)}_{\Omega \ell}}\right)\right|_{\dot{\cal I} } = 0,
$$
which implies $\left.W^{(u \ell)}_{\Omega\, \Omega'}\right|_{\dot{\cal I}} = \left.W^{(u \ell)}_{-\Omega\, -\Omega'}\right|_{\dot{\cal I}} = 0$ and,
for $\Omega \neq \Omega'$,
$\left.W^{(u \ell)}_{\Omega\, -\Omega'}\right|_{\dot{\cal I}} = \left.W^{(u \ell)}_{-\Omega\, \Omega'}\right|_{\dot{\cal I}} = 0$.
Therefore, using the analogous of Eq.~(\ref{Wint}),
$$  
W^{(u\ell)}_{\pm\Omega \,\pm\Omega'} (\rho)
= -i(\pm\Omega\mp\Omega') \int_{\rho_-}^\rho d\rho'\frac{\varepsilon_\bot}{\mu_\bot}
\overline{g^{\rm (TE)}_{\Omega \ell}} {g}^{\rm (TE)}_{\Omega' \ell},
$$
into Eq.~(\ref{AAWuTE}), we finally obtain
\begin{eqnarray}
 \left(A^{\rm ({\it u}TE)}, A'^{\rm ({\it u}TE)}\right) 
 &=&   4\Omega\ell (\ell+1)\delta_{\ell \ell'} \delta_{m m'}\nonumber \\
& & \times {\rm Im}\left(\overline{\alpha_{\Omega \ell}} \beta_{\Omega\ell}\right)
\int_{\cal I} d\varrho\;
\overline{g^{\rm (TE)}_{\Omega \ell}} {g}^{\rm (TE)}_{\Omega' \ell},\nonumber \\
\label{normalizacaossuTE}
\end{eqnarray}
where ${\rm Im}(z)$ stands for the coefficient of the imaginary part of the complex number $z$.
Thus, imposing orthonormality of these modes --- for orthonomalized $ {g}^{\rm (TE)}_{\Omega \ell}$
(in the $L^2({\cal I},d\varrho)$ inner product) ---, the general expression for $\alpha_{\Omega\ell}$ and $\beta_{\Omega\ell}$
(up to rephasing, $\alpha_{\Omega\ell}\mapsto e^{i\delta}\alpha_{\Omega\ell}$, $\beta_{\Omega\ell} \mapsto e^{i\delta} \beta_{\Omega\ell} $,
and time resetting, $\alpha_{\Omega\ell}\mapsto e^{\Omega t_0}\alpha_{\Omega\ell}$, $\beta_{\Omega\ell} \mapsto e^{-\Omega t_0} \beta_{\Omega\ell} $)
read
\begin{eqnarray}
\alpha_{\Omega \ell} &=& \frac{e^{-i\kappa/2}}{2\sqrt{\Omega \ell (\ell+1) \sin \kappa}},
\label{alphauTE}
\\
 \beta_{\Omega \ell} &=& \frac{e^{i\kappa/2}}{2\sqrt{\Omega \ell (\ell+1) \sin \kappa}},
\label{betauTE}
\end{eqnarray}
with $0<\kappa<\pi$ being an arbitrary constant.

\subsection{TM modes}
\label{subsec:TM}

\subsubsection{Stable}
\label{subsubsec:sTM}

Now, substituting ${\bf A}^{\rm (TM)} = (r^{-2} \varepsilon_\parallel^{-1}\Delta^{\!(0)}_S,\partial_\theta
\partial_\varrho,
\partial_\varphi\partial_\varrho)\phi$ into
 Eq.~(\ref{sesqss}), with $\phi = e^{-i\omega\tau}Y_{\ell m}(\theta,\varphi)f_{\omega \ell}^{\rm (TM)}(r)$, 
 and evaluating the angular integrals (similarly to the previous TE case), we obtain:
 \begin{widetext}
\begin{eqnarray}
 \left(A^{\rm (TM)}, A'^{\rm (TM)}\right) &=&\ell (\ell+1)
 \delta_{\ell \ell'}\delta_{m m'}
\int_{\cal I}d\varrho\,e^{i(\omega-\omega')\tau}
\left\{
(\omega+\omega')\left[\frac{d}{d\varrho}\overline{f^{\rm (TM)}_{\omega \ell}}
\frac{d}{d\varrho} {f}^{\rm (TM)}_{\omega' \ell}+\frac{\gamma^2(1-n_\parallel^2 v^2)\ell (\ell+1)}
{\varepsilon_\parallel \varepsilon_\bot r^2}
\overline{f^{\rm (TM)}_{\omega \ell}}
 {f}^{\rm (TM)}_{\omega' \ell}
\right]\right.
\nonumber \\
& & \hskip 5.0cm\left.
+\frac{i \gamma^2(n_\parallel^2-1) v}{\varepsilon_\bot}\left(\omega^2
\overline{f^{\rm (TM)}_{\omega \ell}} \frac{d}{d\varrho}{f}^{\rm (TM)}_{\omega' \ell}-\omega'^2
{f}^{\rm (TM)}_{\omega' \ell}\frac{d}{d\varrho}\overline{f^{\rm (TM)}_{\omega \ell}} 
\right)
\right\},
\label{AApTM}
\end{eqnarray}
\end{widetext}
where recall that $dr/d\varrho = \gamma^2(1-n_\parallel^2 v^2)/\varepsilon_\bot$. The strategy to simplify the
expression above is the same applied in the TE case. Define
\begin{eqnarray}
{\cal W}^{(\ell)}_{\omega \omega'} &:=&\frac{1}{(\omega-\omega')}
\nonumber \\
& & \times
\left(\omega^2
\overline{f^{\rm (TM)}_{\omega \ell}} \frac{d}{d\varrho}{f}^{\rm (TM)}_{\omega' \ell}-\omega'^2
{f}^{\rm (TM)}_{\omega' \ell}\frac{d}{d\varrho}\overline{f^{\rm (TM)}_{\omega \ell}} 
\right).\nonumber \\
\label{calWdf}
\end{eqnarray}
One can easily check, using Eq.~(\ref{fTMss}), that 
\begin{eqnarray}
\frac{d}{d\varrho}{\cal W}^{(\ell)}_{\omega \omega'} &=&(\omega+\omega')
\left[\frac{d}{d\varrho}\overline{f^{\rm (TM)}_{\omega \ell}}
\frac{d}{d\varrho} {f}^{\rm (TM)}_{\omega' \ell}\right. 
\nonumber \\
& & \left.+\frac{\gamma^2(1-n_\parallel^2 v^2)\ell (\ell+1)}{\varepsilon_\parallel \varepsilon_\bot r^2}
\overline{f^{\rm (TM)}_{\omega \ell}}
 {f}^{\rm (TM)}_{\omega' \ell}
\right].\;\;\;\;\;\;
 \label{dcalWdrho}
\end{eqnarray}
Therefore, we can put Eq.~(\ref{AApTM}) in the same form as Eq.~(\ref{AAWTE}), with 
$W\mapsto{\cal W}$ and $\rho\mapsto \varrho$. Now, orthogonality of the modes demand
that $f_{\omega\ell}^{\rm (TM)}$ satisfy either Dirichlet or Neumann boundary conditions
at $\dot{\cal I}$, which leads to 
\begin{eqnarray}
& & \left(A^{\rm (TM)}, A'^{\rm (TM)}\right) =
 \ell (\ell+1)\delta_{\ell \ell'} \delta_{m m'} 
\left.\left[{\cal W}^{(\ell)}_{\omega \omega'}\right]\right|_{\dot{\cal I}},\;\;\;\;\;\;
\label{AAWTM}
\end{eqnarray}
In order to simplify even further the expression above,
note that using again Eq.~(\ref{fTMss}) in Eq.~(\ref{dcalWdrho}), we can write
\begin{eqnarray}
\frac{d}{d\varrho}{\cal W}^{(\ell)}_{\omega \omega'} &=&\frac{(\omega+\omega')}{2}\nonumber \\
& & \times
\left[
\frac{d^2}{d\varrho^2}+\frac{\mu_\bot}{\varepsilon_\bot}(\omega^2+\omega'^2)\right]\left(\overline{f^{\rm (TM)}_{\omega \ell}}
 {f}^{\rm (TM)}_{\omega' \ell}\right),\nonumber \\
\label{dcalWdrho2}
\end{eqnarray}
whose integration on ${\cal I}$ gives us $\left.\left[{\cal W}^{(\ell)}_{\omega \omega'}\right]\right|_{\dot{\cal I}}$,
which substituted into Eq.~(\ref{AAWTM}) finally leads to
\begin{eqnarray}
 \left(A^{\rm (TM)}, A'^{\rm (TM)}\right)& =&  2\omega^3
\ell (\ell+1) \delta_{\ell \ell'} \delta_{m m'}
\nonumber \\
& & \times
 \int_{\cal I}
d\rho \,
\overline{f^{\rm (TM)}_{\omega \ell}} {f}^{\rm (TM)}_{\omega' \ell}.
\label{normalizacaossTM}
\end{eqnarray}
(Notice that the integration variable is $\rho$.)

\subsubsection{Unstable}
\label{subsubsec:uTM}

Generic unstable TM modes are given by 
${\bf A}^{\rm ({\it u}TM)} =(r^{-2} \varepsilon_\parallel^{-1}\Delta^{\!(0)}_S,\partial_\theta
\partial_\varrho,
\partial_\varphi\partial_\varrho)\phi$  with
\begin{eqnarray}
\phi =(\alpha_{\Omega \ell} e^{\Omega\tau} + \beta_{\Omega \ell} e^{-\Omega\tau} )Y_{\ell m}(\theta,\varphi)g_{\Omega \ell}^{\rm (TM)}(r),
\label{generalUTM}
\end{eqnarray}
where $\alpha_{\Omega \ell}$ and $\beta_{\Omega \ell}$ are complex constants and $g_{\Omega \ell}^{\rm (TM)}(r)$ is a solution of Eq.~(\ref{fTMss}) with $\omega^2 = -\Omega^2$
($\Omega >0$, without loss of generality) satisfying Dirichlet or Neumann boundary conditions. 
Once more, sesquilinearity of Eq.~(\ref{sesqss}) makes it easy to calculate $ \left(A^{\rm ({\it u}TM)}, A'^{\rm ({\it u}TM)}\right) $ from Eq.~(\ref{AAWTM}) with the substitution
$\omega \mapsto \mp i\Omega$ and $\omega'\mapsto \pm i\Omega'$:
\begin{eqnarray}
\left(A^{\rm ({\it u}TM)}, A'^{\rm ({\it u}TM)}\right) 
&=& \ell (\ell+1)\delta_{\ell \ell'} \delta_{m m'} 
\nonumber \\
& &   \!\!\!\!\!\!\!\!\!\!\! \!\!\!\!\!\!\!\!\!\!\!\!\!\!\!\!\!\!\!\!\!\!\!\!\!\!\!\!\!\!\!\!\! \times \left.\left[\overline{\alpha_{\Omega \ell}} \beta_{\Omega'\ell}
{\cal W}^{(u \ell)}_{\Omega\;- \Omega'}+\overline{\beta_{\Omega \ell}} \alpha_{\Omega'\ell}
{\cal W}^{(u \ell)}_{-\Omega \Omega'}\right]\right|_{\dot{\cal I}},\;\;\;
\label{AAWuTM}
\end{eqnarray}
where
\begin{eqnarray}
{\cal W}^{(u\ell)}_{\pm\Omega \,\pm\Omega'} &:=&\frac{-i}{(\pm\Omega\pm\Omega')}\left(
\Omega^2
\overline{g^{\rm (TM)}_{\Omega \ell}} \frac{d}{d\rho}{g}^{\rm (TM)}_{\Omega' \ell}
\right.\nonumber \\
& & \left.
-
\Omega'^2
{g}^{\rm (TM)}_{\Omega' \ell}\frac{d}{d\rho}\overline{g^{\rm (TM)}_{\Omega \ell}}\right)
\nonumber \\
&=& \frac{i(\pm\Omega\mp\Omega')}{2}\left[-\frac{d}{d\varrho}\left(\overline{g^{\rm (TM)}_{\Omega \ell}} {g}^{\rm (TM)}_{\Omega' \ell}\right)\right.\nonumber \\
& &\left.+ (\Omega^2 +\Omega'^2) \int_{\varrho_-}^\varrho d\varrho'\frac{\mu_\bot}{\varepsilon_\bot}
\overline{g^{\rm (TM)}_{\Omega \ell}} {g}^{\rm (TM)}_{\Omega' \ell}\right].\;\;\;\;\;\;
\label{calWudf}
\end{eqnarray}
[In the last passage of the expression above we used the analogous of Eq.~(\ref{dcalWdrho2}).]
\begin{eqnarray}
 \left(A^{\rm ({\it u}TM)}, A'^{\rm ({\it u}TM)}\right) 
 &=& -  4\Omega^3\ell (\ell+1)\delta_{\ell \ell'} \delta_{m m'}\nonumber \\
& & \times {\rm Im}\left(\overline{\alpha_{\Omega \ell}} \beta_{\Omega\ell}\right)
\int_{\cal I} d\rho\;
\overline{g^{\rm (TM)}_{\Omega \ell}} {g}^{\rm (TM)}_{\Omega' \ell},\nonumber \\
\label{normalizacaossuTM}
\end{eqnarray}
Imposing orthonormality of these modes --- for orthonomalized $ {g}^{\rm (TM)}_{\Omega \ell}$
(in the $L^2({\cal I},d\rho)$ inner product) ---, the general expression for $\alpha_{\Omega\ell}$ and $\beta_{\Omega\ell}$
(again, up to rephasing
and time resetting)
can be expressed as
\begin{eqnarray}
\alpha_{\Omega \ell} &=& \frac{e^{i\kappa/2}}{2\sqrt{\Omega^3 \ell (\ell+1) \sin \kappa}},
\label{alphauTM}
\\
 \beta_{\Omega \ell} &=& \frac{e^{-i\kappa/2}}{2\sqrt{\Omega^3 \ell (\ell+1) \sin \kappa}},
\label{betauTM}
\end{eqnarray}
with $0<\kappa<\pi$.



\begin{thebibliography}{99}
\bibitem{Foot1}
By ``spatial'' here we mean that the contraction of any index of these tensors with the medium's
four-velocity $v^a$ vanishes.
\end{thebibliography}
\end{document}